\documentclass{aa}  

\usepackage{graphicx}
\usepackage{txfonts}
\usepackage[]{hyperref}
\usepackage{amsmath}
\usepackage{placeins}
\usepackage{float}

\begin{document}

   \title{The velocity dispersion function of red galaxies in four \textit{Hubble} Frontier Fields galaxy clusters}

\author{G.~Granata\inst{\ref{unife},\ref{unimi},\ref{icg}}, 
        L.~Tortorelli\inst{\ref{lmu}},
        C.~Grillo\inst{\ref{unimi},\ref{iasf}},
        P.~Rosati\inst{\ref{unife},\ref{inafoas}},
        M.~D'Addona\inst{\ref{unisa},\ref{inafna},\ref{unife}},
        A.~Mercurio\inst{\ref{unisa},\ref{inafna},\ref{infnsa}},
        G.~Angora\inst{\ref{inafna},\ref{unife}},
        P.~Bergamini\inst{\ref{inafoas}},
        \and
        G.~B.~Caminha\inst{\ref{mpa},\ref{tum}}
}
\date{Received 2 December 2025 / Accepted 19 March 2026}
\institute{
    Dipartimento di Fisica e Scienze della Terra, Universit\`a degli Studi di Ferrara, via Saragat 1, 44122 Ferrara, Italy \label{unife}
    \and
    Dipartimento di Fisica, Universit\`a  degli Studi di Milano, via Celoria 16, 20133 Milano, Italy \label{unimi}    
    \and
    Institute of Cosmology and Gravitation, University of Portsmouth, Burnaby Rd, Portsmouth PO1 3FX, UK \label{icg} \\
    e-mail: \href{mailto:giovanni.granata@port.ac.uk}{\tt giovanni.granata@port.ac.uk}
    \and
    Universitäts-Sternwarte, Fakultät für Physik, Ludwig-Maximilians-Universität München, Scheinerstraße 1, 81679 München, Germany\label{lmu}
    \and
    INAF -- IASF Milano, via Corti 12, 20133 Milano, Italy \label{iasf}
    \and
    INAF -- OAS, Osservatorio di Astrofisica e Scienza dello Spazio di Bologna, via Gobetti 93/3, 40129 Bologna, Italy \label{inafoas}
    \and
    Dipartimento di Fisica ``E.R. Caianiello'', Università Degli Studi di Salerno, Via Giovanni Paolo II, 84084 Fisciano (SA), Italy \label{unisa}
    \and
    INAF -- Osservatorio Astronomico di Capodimonte, Salita Moiariello 16, 80131 Napoli, Italy \label{inafna}
    \and
    INFN -- Gruppo Collegato di Salerno – Sezione di Napoli, Dipartimento di Fisica ``E.R. Caianiello'', Università di Salerno, Via Giovanni Paolo II, 84084 Fisciano (SA), Italy \label{infnsa}
    \and
    Max-Planck-Institut f{\"u}r Astrophysik, Karl-Schwarzschild Stra{\ss}e 1, 85748 Garching, Germany \label{mpa}
    \and
    Technical University of Munich, TUM School of Natural Sciences, Physics Department,  James-Franck-Stra{\ss}e 1, 85748 Garching, Germany \label{tum}
}

\authorrunning{G. Granata et al.}
% \abstract{}{}{}{}{} 
% 5 {} token are mandatory
 
  \abstract{We present a detailed study of the stellar kinematic properties of red member galaxies in the cores of four strong lensing galaxy clusters at intermediate redshifts included in the \textit{Hubble} Frontier Fields programme: Abell 2744 ($z=0.307$), Abell S1063 ($z=0.346$), MACS J0416.1$-$2403 ($z=0.397$), and MACS J1149.6$+$2223 ($z=0.542$). We focussed on a large sample of 723 red cluster members in the four clusters, selected spectroscopically and photometrically, and we measured their structural parameters using \texttt{MORPHOFIT} for all \textit{Hubble} Frontier Fields bands. Taking advantage of deep (3.1~h to 17~h of exposure time) integral-field spectroscopic data from the Multi Unit Spectroscopic Explorer (MUSE) on the Very Large Telescope, available for the cores of all four clusters, we tested a pipeline based on the public spectral fitting code \texttt{pPXF} to reliably and systematically measure the line-of-sight stellar velocity dispersion $\sigma$ of cluster members with a spectral $S/N\geq 10$, with a statistical uncertainty consistently below 5\%. The resulting catalogue contains 213 measured $\sigma$ values across the four clusters. Combining stellar kinematics and multi-band galaxy morphology, we calibrated the Fundamental Plane relation in the rest-frame $r$ band for the early-type cluster members, selected from their colour and morphology; we found compatible parameters both across the clusters and in comparison with large samples of early-type field galaxies, and noted hints of zero-point evolution with redshift. Finally, we used the calibrated Fundamental Plane relations to assign a velocity dispersion value to all 723 red cluster members and studied the velocity dispersion function for each cluster, down to $\log \sigma \, \mathrm{[km \, s^{-1}] = 1.5}$. In spite of the intrinsic variability between the four clusters resulting from their assembly history, a Schechter function fit of the velocity functions suggests compatible parameters: a positive $\alpha$ slope with values in the range $0.55-1.60$ and $\log\sigma^*\, [\mathrm{km\,s^{-1}}]$ between $2.18$ and $2.47$. Unlike previous works, we extended the systematic study of the central velocity dispersion of cluster galaxies to lower-$\sigma$ regimes. We suggest that deeper insights on the evolution of member galaxies may be obtained with a larger cluster sample spanning a wider redshift range.
    }
   \keywords{galaxies: elliptical and lenticular, cD -- galaxies: clusters: individual: Abell 2744 --
galaxies: clusters: individual: Abell S1063 -- galaxies: clusters: individual: MACS J0416.1–2403 --
galaxies: clusters: individual: MACS J1149.6+2223 -- galaxies: kinematics and dynamics
               }

   \maketitle

\section{Introduction}\label{s1}

Galaxy clusters serve as a crucial link between astrophysics and cosmology, acting as key tracers of structure formation through hierarchical assembly and as laboratories for studying galaxy evolution in dense environments. As smaller haloes are accreted onto larger structures during cluster formation, they undergo a range of physical processes \citep[][]{kravtsov12}. Tidal stripping and dynamical heating can remove substantial mass, while baryonic processes, such as radiative cooling and feedback from star formation and active galactic nuclei, reshape their internal structure and regulate their star formation activity. Consequently, the dark matter (DM) distribution in clusters retains signatures of both the large-scale assembly history and the micro-physical interactions between DM and baryons \citep{giocoli08,giocoli10}. 

Strong lensing has emerged as a prime technique to map the total mass distribution of massive galaxy clusters. When combined with observational probes of the mass of baryons, it lets us investigate the physical properties and mass structure of DM haloes \citep[e.g.][]{natarajan97,caminha19,granata22,granata23}. However, several studies have shown that the recovery of the mass distribution of galaxy-scale DM haloes (or sub-haloes) is hindered by parametric degeneracies whose impact can be significantly reduced with independent observational priors obtained from the stellar kinematics of the cluster members \citep[][]{bergamini19,granata22}. The stellar line-of-sight velocity distribution of early-type galaxies (ETGs) is a direct and relatively unbiased tracer of the total gravitational potential, less affected by systematics or a priori assumptions than luminosity- or stellar-mass-based estimates. The central stellar velocity dispersion (hereafter indicated with $\sigma$), in particular, serves as a robust probe of the gravitational potential well in ETGs. Its distribution, or function, for a galaxy population offers a way to link its properties with theoretical models of galaxy formation and evolution.

The velocity dispersion function (VDF) of ETGs has been extensively characterised thanks to large spectroscopic surveys such as the Sloan Digital Sky Survey (SDSS) and the SDSS Baryon Oscillation Spectroscopic Survey (BOSS; e.g. \citealt{sheth03,mitchell05, choi07, bernardi10, monterodorta17, sohn17a}), and in the NMBS COSMOS and UKIDSS UDS deep fields  \citep{bezanson12}. These efforts have revealed the dependence of the velocity dispersion function on galaxy morphology, environment, and redshift. Following \citet{sheth03}, the VDF of red galaxies is typically modelled by a modified Schechter function:
\begin{equation}\label{modsch}
\phi(\sigma)\, \mathrm{d}\sigma = \phi^* \left(\frac{\sigma}{\sigma^*}\right)^\alpha 
\frac{\exp\left[- \left(\frac{\sigma}{\sigma^*}\right)^\beta \right]}{\Gamma(\alpha/\beta)} 
\, \beta\, \frac{\mathrm{d}\sigma}{\sigma}.
\end{equation}
Here $\phi^*$ is the normalisation, $\sigma^*$ a characteristic velocity dispersion value, and $\alpha$, $\beta$ control the low- and high-$\sigma$ slopes. This form originates from the connection between galaxy luminosity and velocity dispersion via the Faber-Jackson relation \citep{faber76} and reflects the broader success of the Schechter formalism in describing galaxy demographics \citep{schechter76}. Variations in the parameters derived between studies arise primarily from the differing selection criteria for ETGs and sample completeness. Generally, these functions exhibit a downturn at $\log \sigma [\mathrm{km\,s^{-1}}] < 2.35$, governed by positive $\alpha$ values. Notably, \citet{bezanson12} find minimal evolution at the high-$\sigma$ end, suggesting that massive quiescent galaxies formed early, while the increase at lower $\sigma$ reflects ongoing quenching in intermediate-mass systems.

Measuring the VDF of cluster members, however, requires spatially dense spectroscopic surveys and very deep observations to achieve the necessary spectral signal-to-noise ratio ($S/N$) for a large, complete, and unbiased samples of member galaxies. \citet{sohn17} calibrated the VDF for the quiescent galaxies in Coma and Abell 2029  using data from the wide-field spectrograph MMT/Hectospec and found clear deviations from the field VDF. In particular, the cluster VDF is flatter at low dispersions, though this difference may reflect incompleteness of the field sample \citep{sohn17}. An excess at $\log \sigma[\mathrm{km\,s^{-1}}] > 2.4$ is also seen, likely due to the presence of BCGs in cluster cores. \citet{sohn20} extended these results to nine strong lensing clusters at $0.18 < z < 0.29$, using data from the same instrument, and confirming both the high-$\sigma$ excess and compatibility with Coma and Abell 2029, down to $\log \sigma [\mathrm{km\,s^{-1}}] \sim 2.05$.

Our aim is to  extend the study of the kinematic properties of red cluster members to higher-redshift and lower-mass galaxies by taking advantage of deep observations from the integral-field Multi Unit Spectroscopic Explorer \citep[MUSE,][]{bacon12} at the Very Large Telescope (VLT), which has allowed systematic spectroscopic coverage of the cores of strong lensing galaxy clusters, with exposure times occasionally above $10 \, \rm h$. The depth of the observations extends kinematic measurements to fainter cluster members \citep[see][]{bergamini21}, and the integral-field nature of the instrument mitigates the impact of the high galactic density in cluster cores. For this purpose, we have developed a pipeline to systematically measure the central stellar velocity dispersion values of the largest possible set of cluster members observed by MUSE. The relatively small MUSE field of view (FoV) of $1'\times1'$ limits the region explored to the cluster core (i.e. a few hundred kiloparsecs from the cluster centre). However, we can calibrate scaling laws, such as the Fundamental Plane relation \citep[][hereafter FP]{dressler87,djorgovski87,bender92}, that link the kinematic properties of cluster members to their structural parameters, and then use these laws to study the kinematic properties of all members within the larger FoV of the photometric observations available. We measure the stellar VDF of red cluster members and discuss the results in terms of the observational picture of galaxy formation and evolution in cluster cores.

The paper is organised as follows. In Sect. \ref{s2} we give details on the sample of four galaxy clusters studied and the observational dataset on which this work is based. In Sect. \ref{s3} we present our velocity dispersion measurement pipeline, and the catalogue of velocity dispersion values we measured. In Sect. \ref{s4} we calibrate the FP for the ETGs in the four clusters. In Sect. \ref{s5} we use them to extend the kinematic study to all red galaxies for which we have measured structural parameters, and we calibrate the VDF for the four clusters. Finally, in Sect. \ref{s6} we summarise our results. Throughout this work, we use a flat $\mathrm{\Lambda}$CDM cosmology with $\Omega_\mathrm{m}=0.3$ and $H_0 = 70$ $\mathrm{km \, s^{-1} \, Mpc^{-1}}$.

\section{The galaxy cluster sample}\label{s2}

We selected a sample of four massive strong lensing galaxy clusters at $0.31<z<0.54$ with deep, homogeneous photometric and spectroscopic coverage, and a state-of-the-art strong lensing model available: Abell 2744 \citep[$z=0.307$, hereafter A2744,][]{allen98,ebeling10}, Abell S1063 \citep[$z=0.346$, hereafter AS1063,][]{abell89,karman15,mercurio21}, MACS J0416.1$-$2403 \citep[$z=0.397$, hereafter M0416,][]{ebeling01,balestra16}, and MACS J1149.6$+$2223 \citep[$z=0.542$, hereafter M1149,][]{smith09,grillo16}.

Optical and near-infrared imaging for the clusters was guaranteed by the \textit{Hubble} Space Telescope (HST). All four clusters were part of the \textit{Hubble} Frontier Fields \citep[HFF,][]{lotz17} programme, and targeted with deep exposures (140 HST orbits) in seven ACS and WFC3 filters (F435W, F606W, F814W, F105W, F125W, F140W, and F160W). The HFF data products were reduced with the HST science data products pipeline. In Fig. \ref{fig:observations}, we show the layout of the four clusters with RGB images obtained from the HFF observations.
%dark matter mass distributions of \textit{Hubble} treasury clusters and the foundations of LCDM structure formation models,
The cores of the four clusters were covered with MUSE, within its $1'\times1'$ FoV, with a $0.2''\times0.2''$ pixel-scale, a spectral range between $4650 \, \rm \AA$ and $9300 \, \rm \AA$, a spectral resolution between $R \approx 1750$ at $4650 \, \rm \AA$ and $R \approx 3750$ at $9300 \, \rm \AA$, and a spectral sampling of $1.25 \, \mathrm{\AA/pixel}$. In our work, MUSE data was reduced with the procedure presented in \cite{caminha17a,caminha17b,caminha19}, based on the standard ESO reduction pipeline \citep{weilbacher20}, complemented with the Zurich Atmosphere Purge \citep[ZAP,][]{soto16} sky subtraction procedure. The MUSE footprint within the four cluster cores is shown in red in Fig. \ref{fig:observations}. Further spectroscopic follow-up campaigns were carried out with multi-object spectrographs, such as the CLASH-VLT programme \citep[][]{rosati14} with the VIsible MultiObject Spectrograph (VIMOS) on the VLT, which allowed the measurement of several thousand spectroscopic redshifts within a $20'$ field centred on several strong lensing clusters, including AS1063 and M0416. In the following sub-sections, we provide some further details on the spectroscopic dataset on which this work is based, which was crucial for the construction of state-of-the-art strong lensing models for the four clusters, based on extremely large sets of spectroscopically confirmed multiple images \citep[see e.g.][]{richard21,caminha16,caminha17a,grillo16,schuldt24}, and on a detailed description of the mass structure of the cluster members enhanced by their observed kinematics \citep[see][]{bergamini19,bergamini22,bergamini23,granata22}.

\subsection{Abell 2744}

The MUSE pointings of A2744 used in this work were obtained within the GTO Programme 094.A-0115 (P.I. Richard). The four main quadrants were covered with exposure times between 3.5h and 5h, and a further 2h pointing was obtained for the cluster central region. The detailed exposure times are presented in \cite{mahler18}, where the authors measured a point-spread-function (PSF) full width at half maximum (FWHM) of $0.58''$ for the combined MUSE image. The spectroscopic redshift catalogue for the field of A2744 was completed with ancillary observations. Specifically, A2744 was targeted for 4.4h with VIMOS \citep[PI: Böhringer, spectroscopic catalogue in][]{braglia09}, with
the AAOmega multi-object spectrograph on the AngloAustralian Telescope (AAT; spectroscopic catalogue in \citealt{owers11}), and with the Grism LensAmplified Survey from Space survey \citep[GLASS;][]{treu15,schmidt14}.

\subsection{Abell S1063}

The MUSE observations of Abell S1063 are the combination of two programmes: the first covered the SW half of the cluster core (ESO SV Programme 60.A-9345, P.I. Clement \& Caputi), for a total exposure time of 3.1h and  $1.1''$ seeing \citep{karman15}, while the second covered the NE region of the core (ESO Programme  095.A-0653, P.I. Caputi) with an exposure time of 4.8h and $0.9''$ seeing \citep{karman17}. Abell S1063 was also part of the CLASH-VLT sample; the MUSE and VIMOS spectroscopic redshifts were then combined with measurements from \cite{gomez12} and from GLASS \citep{treu16} within the spectroscopic catalogues presented in \cite{mercurio21}, which includes 1234 cluster members.

\subsection{MACS J0416.1$-$2403}

The MUSE observations of MACS J0416 have separately focussed on the NE and SW regions of its core, respectively. The former was targeted by the GTO programme 094.A0115B (P.I. Richard) and the ultra-deep observations 0100.A0763(A) (P.I. Vanzella), for a total integration time of 17.1h in most of the field and a seeing of approximately $0.6''$ \citep{vanzella21}, the deepest MUSE pointing for a strong lensing cluster. The latter was targeted in the 094.A0525(A) programme (P.I. Bauer), for 11h of integration time and $1.0''$ seeing. Similarly to AS1063, M0416 was targeted with VIMOS within the CLASH-VLT programme \citep{balestra16}.

\subsection{MACS J1149.6$+$2223}

The MUSE coverage of the central regions of M1149 was guaranteed by the DDT programme 294.A-5032 (P.I. Grillo), for a total integration time of 6h and seeing of less than $1.1''$ in almost all exposures \citep{grillo16}, complemented by $5.5 \, \mathrm{h}$ of observations of the north-west region of the cluster within the programme 105.20P5 (PI Mercurio), presented in \cite{schuldt24}. Further spectroscopic data was obtained within the GLASS programme \citep{treu16}.

\begin{figure*}[t!]
  \includegraphics[width = \columnwidth]{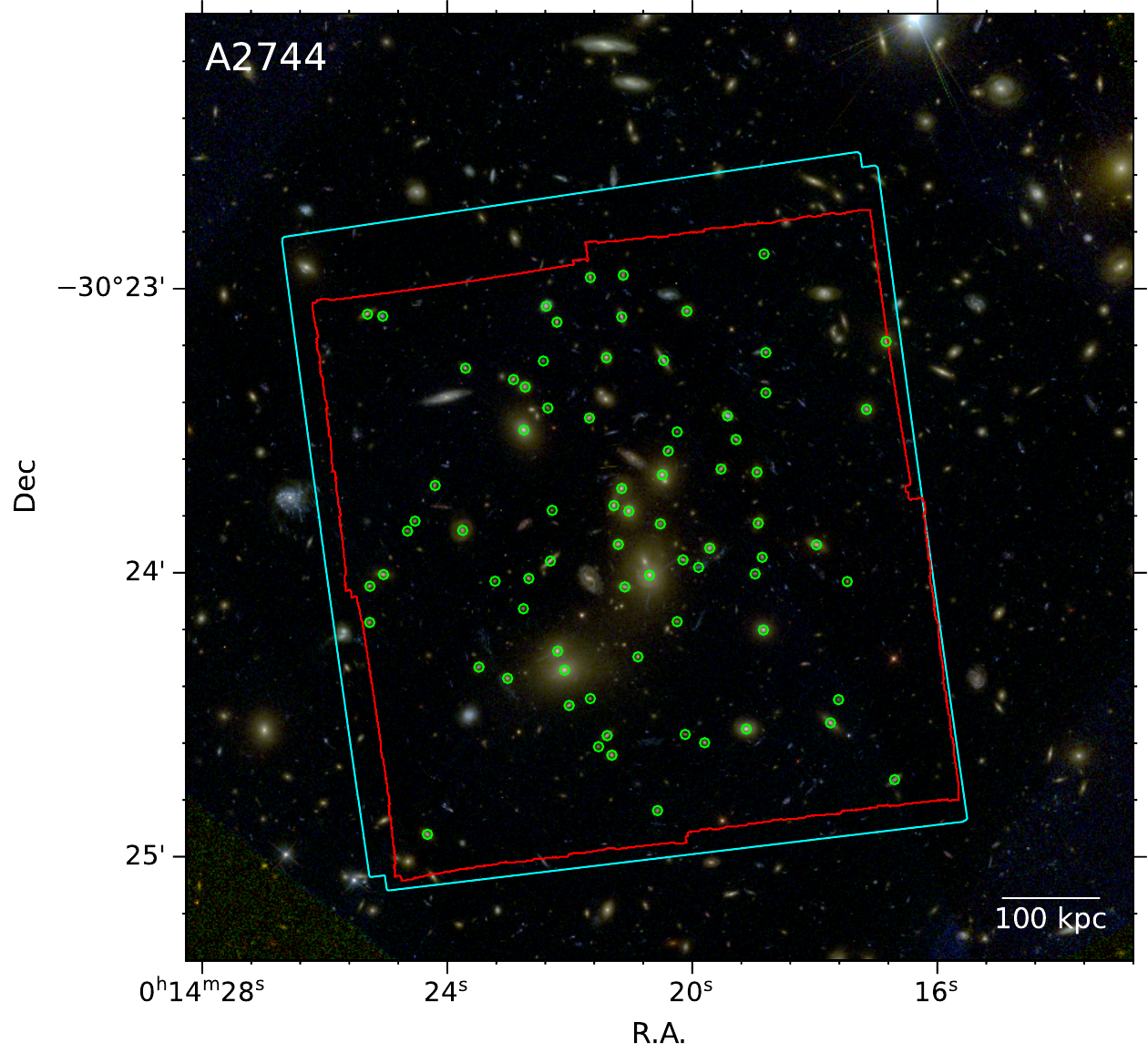}
  \includegraphics[width = \columnwidth]{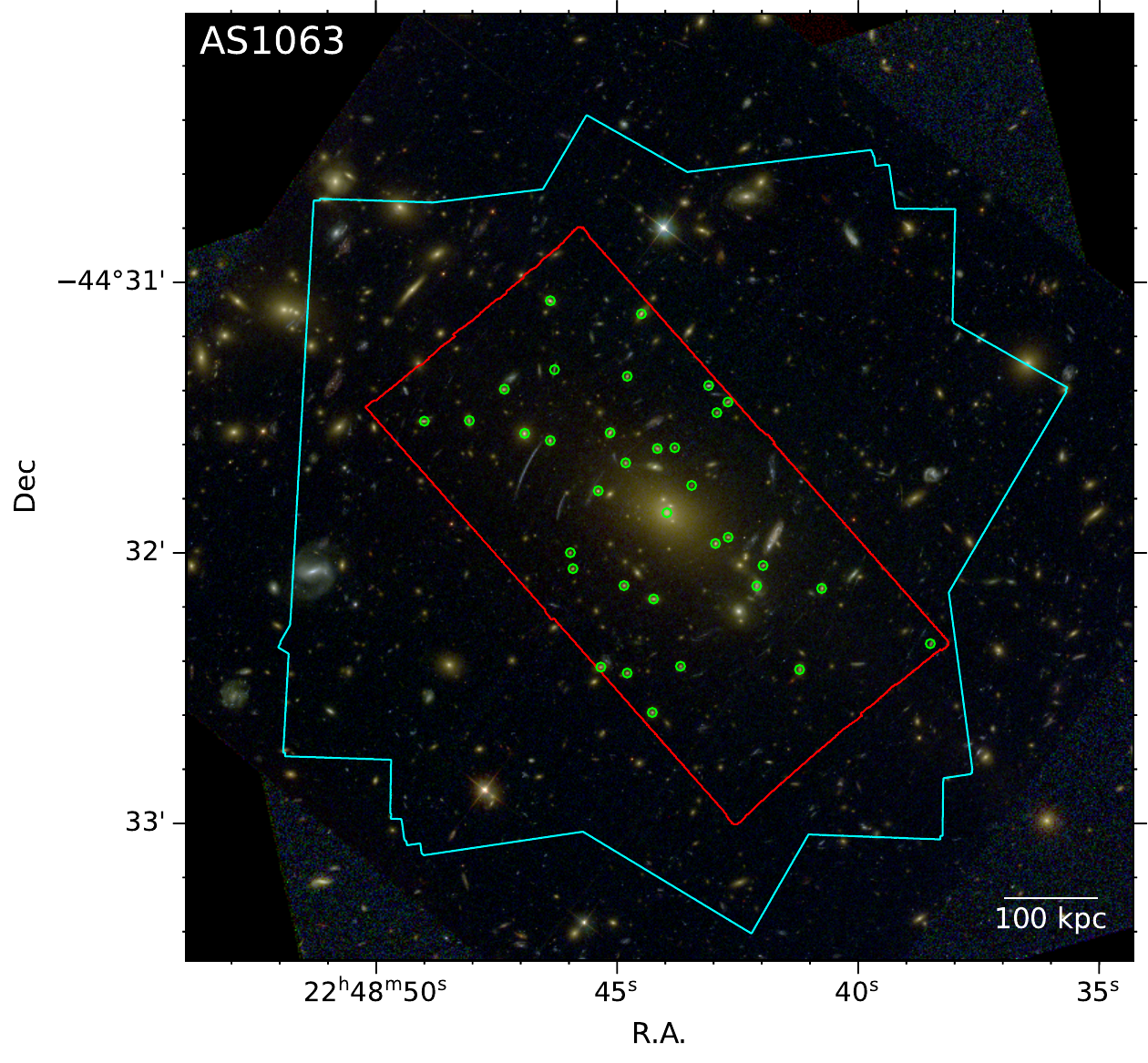}
  \includegraphics[width = \columnwidth]{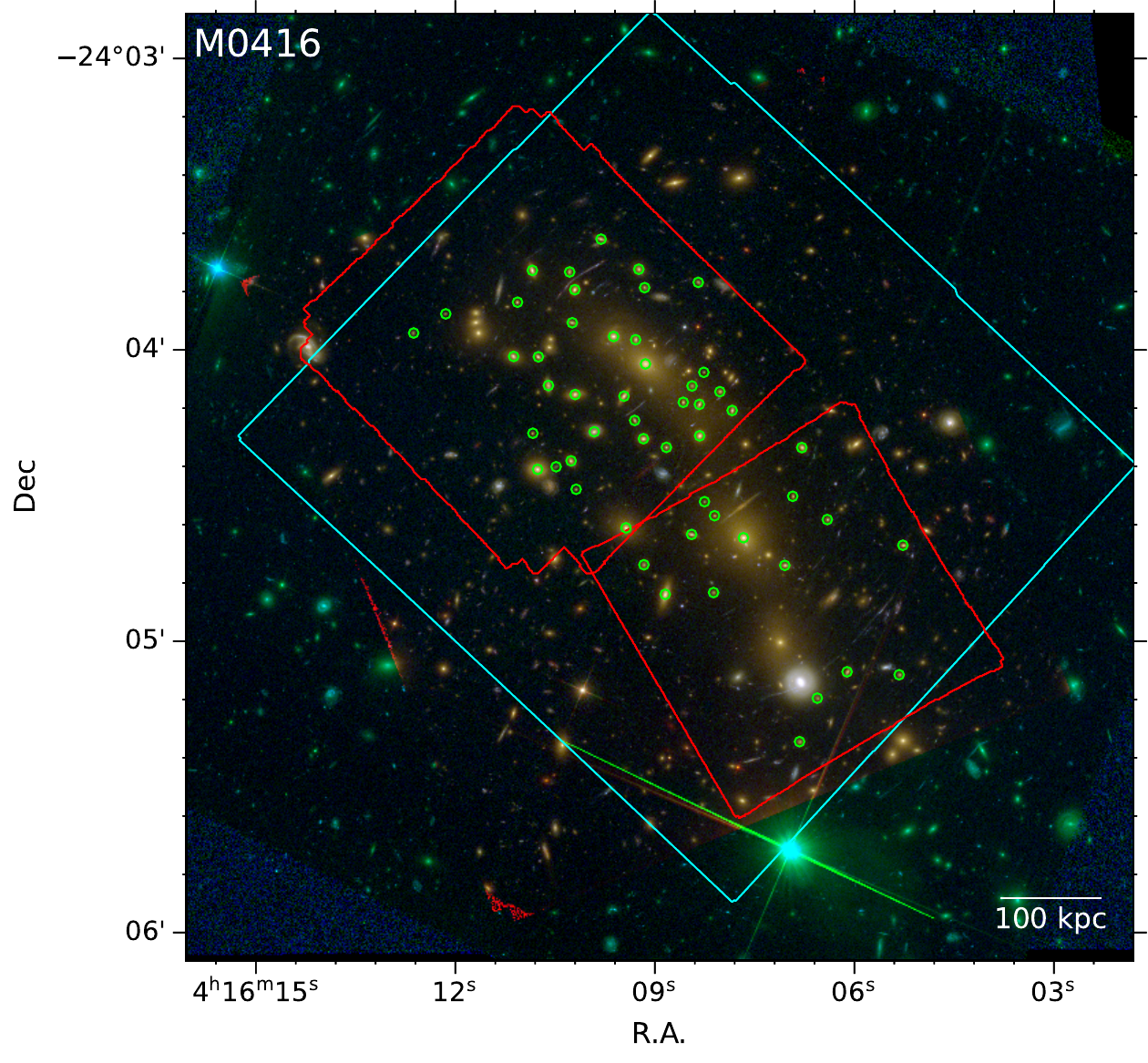}
  \includegraphics[width = \columnwidth]{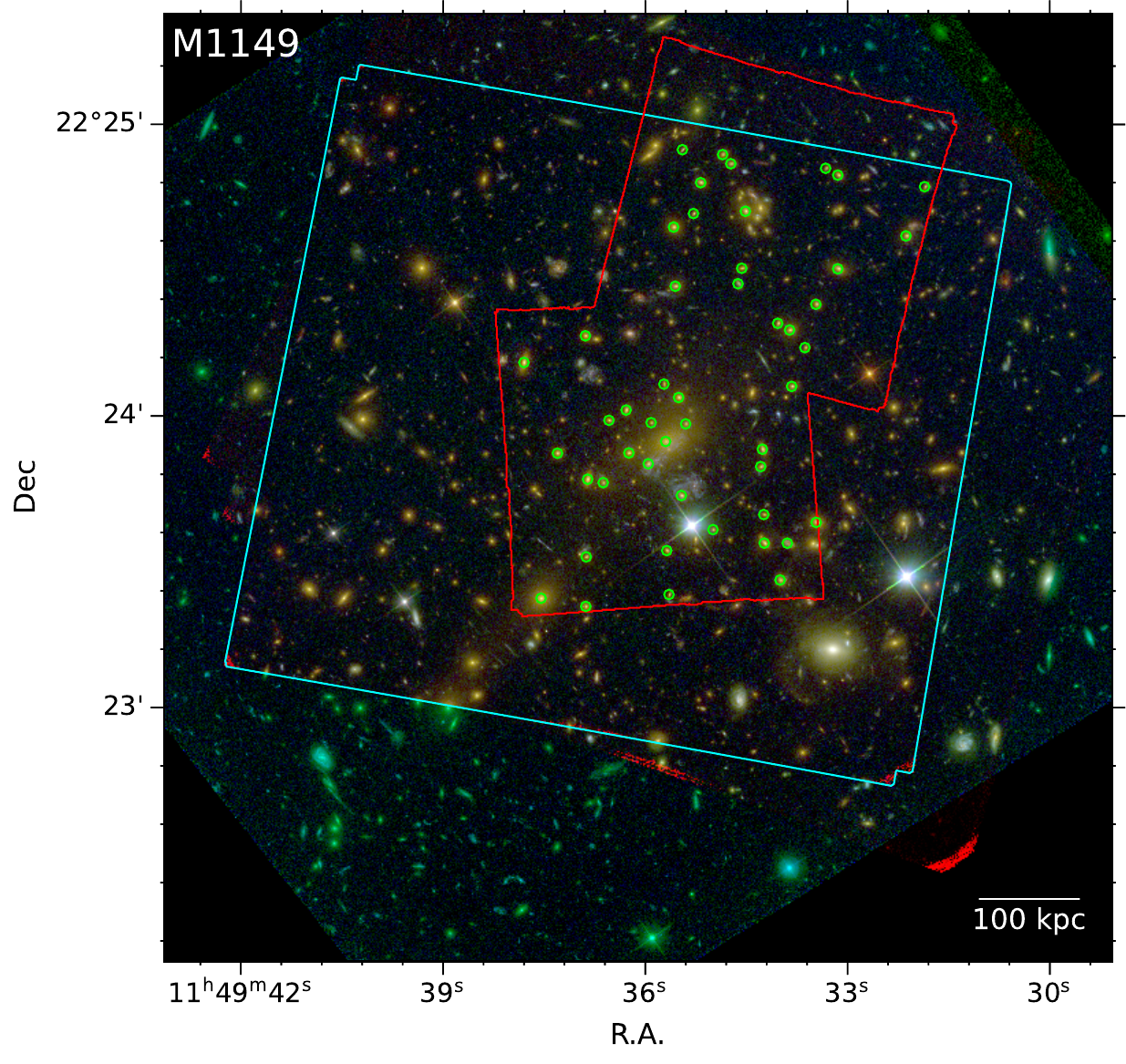}
  \caption{RGB images of the cores of the four clusters from HFF photometry. The red, green, and blue channels were obtained by combining the filters \textit{F105W} + \textit{F125W} + \textit{F140W} + \textit{F160W}, \textit{F606W} + \textit{F814W}, and \textit{F435W}, respectively. We show the FoV of the MUSE observations with red contours, and we mark all the cluster members for which we have measured a velocity dispersion value with green circles. Finally, we show the region intersection of all HFF bands for each cluster with cyan contours.}
  \label{fig:observations}
\end{figure*}

\subsection{The cluster member catalogue}\label{sec:cmcat}

The spectroscopic catalogues presented in the previous section and in the papers therein referenced, built primarily with MUSE and VIMOS data, were the starting point for the selection of the sample of cluster members that we will include in our calibrated VDF. We selected spectroscopic cluster members through analysis of the cluster 1D velocity distribution, following \citet{bergamini23}, \citet{caminha16}, \citet{caminha17a}, and \citet{grillo16} for A2744, AS1063, M0416, and M1149, respectively.  These spectroscopic cluster members catalogues can be reliably extended even to regions not covered by spectroscopy when deep multi-band photometry is available. \cite{angora20} developed a convolutional neural network to identify cluster members on the basis of pan-chromatic imaging alone, achieving a purity-completeness rate larger than $90\%$ down to a HST F814W band magnitude $m_{814}=25$ when all HFF bands are available. We adopted their completed cluster member catalogues, presented in \cite{angora20} for AS1063, M0416, and M1149, and in \cite{bergamini23} for A2744, as the basis for our work, limiting ourselves to the intersection between the footprints of all HFF band for each cluster to ensure uniform completeness of the photometric samples. We show this region in Fig. \ref{fig:observations} with cyan contours for all four clusters. The resulting cluster member catalogues contain 221 (177), 221 (146), 218 (176), and 253 (191) galaxies overall (spectroscopically confirmed) for A2744, AS1063, MACS0416, and MACS1149, respectively.

\section{The velocity dispersion catalogue}\label{s3}

\begin{figure}\label{showfit}
   \centering
   \includegraphics[width=\hsize]{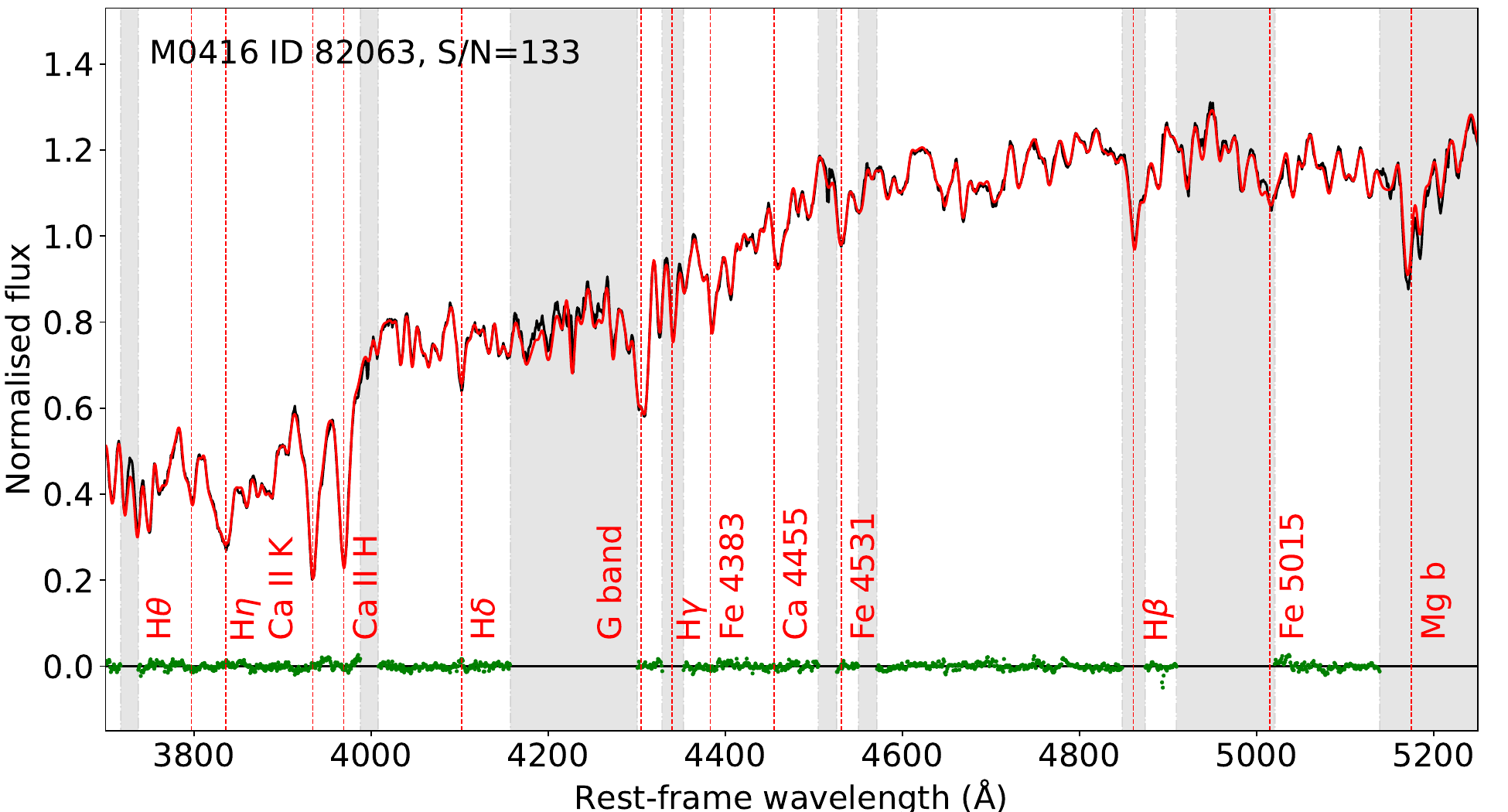}
      \caption{Line-of-sight velocity dispersion fit with \texttt{pPXF} and main spectral absorption features contributing to the measurement. We chose the cluster member ID 82063 of M0416 (R.A.$=64.0401169$, Dec.=$-24.0659218$, $S/N=133$). The regions masked as potentially affected by issues with the subtraction of sky features and emission lines are shaded in grey. We note that the mask in the spectral region around $4200 \, \mathrm{\AA}$ is intended to avoid the effects of the adaptive-optics laser emission used for some MUSE exposures of M0416, and as such was not adopted for the other three clusters. We show the observed spectrum in black, the best fit in red, and the fit-measured spectral residuals in green. The main absorption lines are marked with red vertical dashed lines.}
   \end{figure}

The calibration of the VDF of red cluster members requires a large catalogue of robustly measured velocity dispersion values. Thanks to the deep spatially resolved MUSE spectroscopy, the line-of-sight velocity distribution (LOSVD) of the member galaxies can be inferred from the shift and broadening of their spectral features. Since our main point of interest is the determination of the velocity dispersion of the cluster members (often indicated as $\sigma$), we focussed on constraining the first two moments of the LOSVD, using the public spectral fitting code \texttt{pPXF}\footnote{Version 8.0.2} \citep[penalised pixel fitting][]{cappellari04,cappellari17,cappellari23}. We performed a full-spectrum fit of the continuum and of the absorption lines in the rest-frame wavelength range $[3700-5250]\, \rm \AA$. This spectral range contains several absorption features that can be used to constrain the velocity dispersion with high accuracy for passive members within all four clusters, as shown in Fig. \ref{showfit}. The optimisation simultaneously fits both the stellar population and the LOSVD of the observed galaxy by comparing its spectrum with a combination of stellar templates. These templates were selected from the high-resolution UVB stellar spectra in the X-shooter Spectral Library (XSL) DR3 \citep{verro22} and were convolved with a LOSVD. Our template selection was informed by the work of \cite{knabel25}, who identified template choice as the primary source of systematics in stellar kinematic analyses of ETGs. They reviewed all 830 XSL DR3 spectra, removing those lacking flux calibration or corrections for slit losses and extinction. The remaining templates were then visually inspected, with any suspected issues, such as poor extinction or telluric correction, high noise levels, or the presence of emission lines, flagged along with the affected wavelength regions. The final selection, aimed at ensuring reliable templates on the wavelength range we consider, includes 462 spectra, which we used for our analyses. The fits were performed using a continuum shape correction with 12th-degree additive Legendre polynomials and convolving the spectra with a Gaussian LOSVD, given that we only fitted its first two moments. On the other hand, we did not use any multiplicative polynomial to correct for dust reddening, as we were exclusively focussing on the LOSVD parameters and did not wish to carry out stellar population studies on the spectra. The results of the fit of a high $S/N$ member of M0416 are shown in Fig. \ref{showfit}, as an example. We tested the reliability of the velocity dispersion measurements by building a sample of 16,000 simulated spectra of cluster members, presented in Appendix \ref{appa}. This set-up allows accurate measurements for $S/N>10$ and down to velocity dispersion values of $50 \, \mathrm{km \, s^{-1}}$, comparable with the spectral dispersion of MUSE, as shown in Appendix \ref{appa}. We also used the simulated spectra to obtain a robust estimate of the uncertainty on the measured value of $\sigma$. On one hand, \texttt{pPXF} provides a formal uncertainty on $\sigma$ based on the covariance matrix of the fitted parameters, which we re-scaled under the assumption that the best-fit $\chi^2$ corresponds to the number of degrees of freedom in the fit. While this approach can yield a reasonable order-of-magnitude estimate when the fit quality is high, it tends to significantly underestimate the uncertainty in cases of noisier spectra or low $\sigma$ values \citep{cappellari04}, and in such regimes, simulations offer a more reliable means of assessing the accuracy of the recovered values. In Appendix \ref{appa}, we quantify the relative uncertainty on $\sigma$ as a function of $S/N$, based on the scatter of the recovered $\sigma$ values around the input velocity dispersion for the simulated spectra. Following a similar approach to \cite{bergamini19}, we fitted a relation between the relative uncertainty $\delta \sigma_\mathrm{sim}$ emerging from the simulation and $S/N$, presented in Eq. (\ref{errapp}). As $\delta \sigma_\mathrm{sim}$ goes to zero for $S/N\approx 80$, we disregarded it for $S/N>80$. To be conservative, we combined this simulation-based error in quadrature with the formal uncertainty from \texttt{pPXF} ($\delta \sigma_\texttt{pPXF}$) to derive our final error estimates on $\sigma$:
\begin{equation}\label{errquad}
    \delta \sigma = \sqrt{\left(\delta \sigma_\texttt{pPXF}\right)^2+\left(\delta \sigma_\mathrm{sim}\right)^2}.
\end{equation}

As is clear from the results of our simulations, a sufficient spectral $S/N$ is the most critical aspect to guarantee a reliable velocity dispersion measurement. In order to increase the sample of galaxies with $S/N>10$, we weighted the MUSE cube for the observed surface brightness of the cluster members \citep[as first done by][]{horne86}. We can thus extract the spectra of the member from a larger aperture around their centre, reducing the flux loss that would be given by a smaller aperture, while the brighter pixels, with the highest $S/N$, contribute the most to the overall spectrum. Specifically, we used the HFF surface brightness in the F814W band, given that it corresponds to a wavelength range where several of the absorption features used for the \texttt{pPXF} fit of the galaxy spectra are found, at the redshift of the four clusters. We measured the HFF and MUSE point-spread-functions (PSFs), and degraded the HFF F814W image to the MUSE resolution and pixel-scale, in order to obtain a weight map for the MUSE cube. 

The spectra of all cluster members within the MUSE footprint were extracted summing all pixels within a $1.5''$ radius from their centre of light, weighted as described earlier. To guarantee a reliable velocity dispersion measurement, all spectra with $S/N<10$ (equivalent to $S/N<8.94 \, \mathrm{\AA^{-1}}$, given MUSE spectral sampling of $1.25 \, \mathrm{\AA}$) were removed from the sample. Given the rather large extraction apertures, all members with a source falling within $3''$ from their centre were visually checked and excluded if they showed any significant photometric or spectroscopic contamination from nearby objects. All fits were also visually checked to make sure that the spectral absorption features had been correctly recognised and well fitted. These selections leave us with a catalogue of 76 measured velocity dispersion values out of 176 members with a MUSE spectroscopic redshift for A2744, 34 out of 94 for AS1063, 52 out of 158 for M0416, and 51 out of 128 for M1149. In total, we have a catalogue of 213 measured velocity dispersion values, which we report in Appendix \ref{appc}. The cluster members for which we have a reliable velocity dispersion measurements are shown with green circles in Fig. \ref{fig:observations}.

\begin{figure}
   \centering
   \includegraphics[width=\hsize]{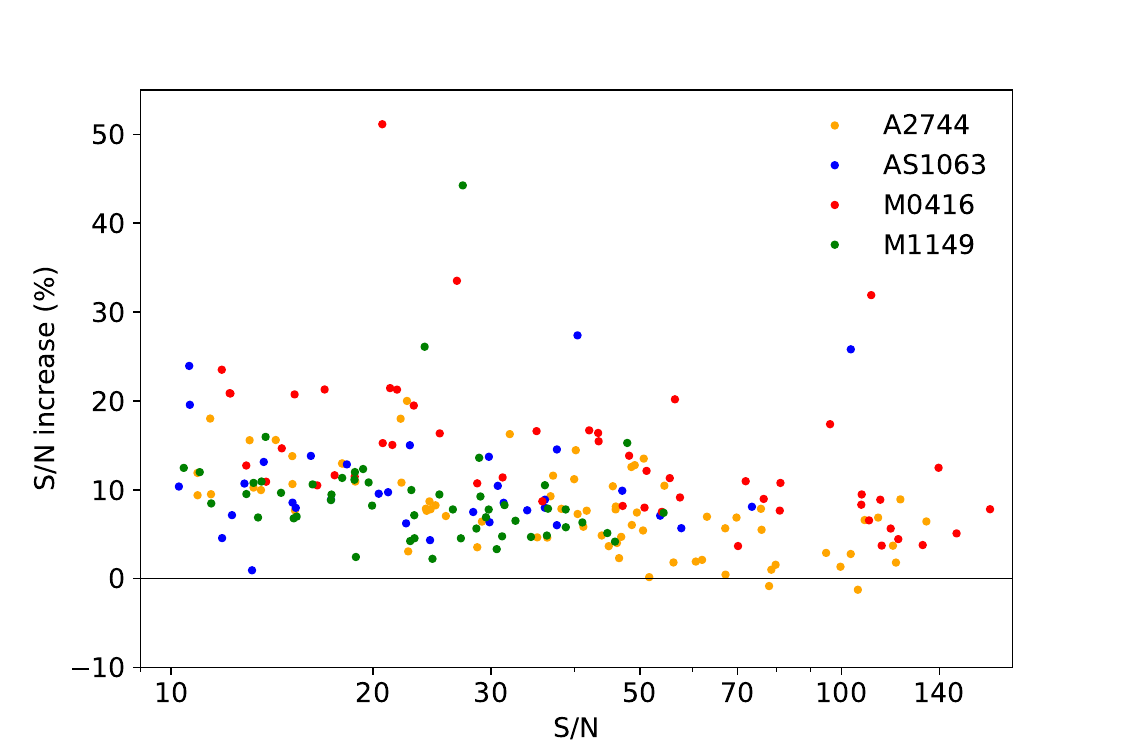}
      \caption{Relative increase in the value of the $S/N$ of the spectra of cluster member galaxies when weighting with their surface brightness in the HST F814W band. We compare the $S/N$ of the spectrum extracted within an aperture of $1.5''$ from the galaxy centre of light from the weighted MUSE cubes with the $S/N$ of the spectrum extracted for the same galaxy within an aperture of $0.8''$ directly from the (non-weighted) MUSE cube.}
         \label{snincrease}
   \end{figure}

The weighting technique for spectral light is crucial in collecting such a large sample, as it allows a significant $S/N$ increase. We compare the $S/N$ of our spectra obtained from the weighted cube with that found extracting them directly from the MUSE cube within an aperture of $0.8''$ from the galaxy centre of light, which has been found to maximise the $S/N$ \citep[see][]{bergamini19,bergamini23}. As is clear from Fig. \ref{snincrease}, our procedure allows a significant increase in the $S/N$, typically in the range $10-30\%$ for $S/N=10-20$, allowing a larger number of objects with $S/N>10$.  

\section{The Fundamental Plane relation for the four clusters}\label{s4}

Observations show that ETGs follow a tight scaling relation linking the kinematic properties, which descend from their total mass distribution, to the luminosity, which depends on the distribution of the stars. This empiric law is known as the FP \citep[][]{dressler87,djorgovski87,bender92}, 
\begin{equation} \label{fp}
    \log R_e = \alpha \log \sigma_0 + \beta \mathrm{SB}_e + \gamma,
\end{equation}
where $R_e$ is the effective radius (i.e. the projected radius enclosing half of the total luminosity), $\sigma_0$ is the central stellar velocity dispersion (i.e. measured within an aperture with a projected radius $R_e/8$, or $\sigma_0= \sigma_\mathrm{ap} \left(R_e/8\right)$), and $\mathrm{SB}_e$ is the average surface brightness within $R_e$. For the calibration of the FP, we complemented the measured velocity dispersion values described in Sect. \ref{s3} with measured structural parameters. This allowed us to leverage the accurate information we obtained on the morphology of all the members towards accurate insights into their kinematics, in spite of the lack of a MUSE spectrum.

We measured the structural parameters for all the members in all HFF bands using the Python package \texttt{MORPHOFIT} \citep{tortorellimercurio,tortorelli18,tortorelli23}, based on \texttt{SExtractor} \citep{bertin96} and \texttt{Galfit} \citep{peng10}, and especially developed to automatically fit the surface-brightness profile of a large number of galaxies in a dense, crowded environment. To properly account for the effect of the intra-cluster light in the cluster core, after an initial \texttt{SExtractor} run to inform the starting points for the light profiles parameters, the code proceeds iteratively in cutouts of increasing size, from stamps around the single galaxies to the full HFF images. All measured structural parameters are available at the CDS and presented in Appendix \ref{appendix:cat_description}. The structural parameters of the cluster members spectroscopically confirmed by MUSE in AS1063, M0416, and M1149 were presented by \citet{tortorelli23}, whilst the remaining have been measured for the first time in this work. In Sect. \ref{s5} our structural parameter catalogue is crucial in order to identify red cluster members and to assign a velocity dispersion value for the object whose spectrum does not have a sufficient $S/N$ for a measurement, and for those that fall outside of the MUSE pointing.

As we wish to compare the FP relations for the four clusters, we need to use magnitude and effective radius measurements in the same rest-frame band. We chose the rest-frame SDSS $r$ band since it is mostly covered, especially in the case of A2744 and AS1063, by the HST F814W band, which guarantees a good $S/N$ down to magnitudes of around 28 and has a small PSF FWHM. Firstly, we performed a $K$ correction of the observed F814W magnitudes to the rest-frame SDSS $r$ band, defined as \citep[e.g.][]{hogg02}
\begin{equation}\label{kcorr}
    K_{814,r} (z)= \frac{1}{1+z} \frac{\int \mathrm{d}\lambda_o \lambda_o L\left(\frac{\lambda_o}{1+z}\right)f_{814}(\lambda_0)}{\int \mathrm{d}\lambda_o \lambda_o g\left(\lambda_o\right)f_{814}(\lambda_0)} \frac{\int \mathrm{d}\lambda_e \lambda_e g(\lambda_e)f_r(\lambda_e)}{\int \mathrm{d}\lambda_e \lambda_e L\left(\lambda_e\right)f_r(\lambda_e)},
\end{equation}
where $z$ is the cluster redshift, $\lambda_o$ and $\lambda_e$ are the observed- and the rest-frame wavelength, $L(\lambda)$ is the galaxy flux as a function of wavelength, and $g(\lambda)$ is the standard source for AB magnitudes. We calibrated the correction using $L(\lambda)$ from a template spectrum built combining ETGs from our MUSE observations. Similarly to \citet{grillo10}, we obtained the effective radius in the SDSS $r$ band as a linear interpolation of those in the HST F814W and F105W bands,
\begin{equation}\label{re}
    R_{e,r}=(1+z)\left(\frac{\lambda_{r}-\lambda_{105}/(1+z)}{\lambda_{814}-\lambda_{105}}R_{e,814}+\frac{\lambda_{r}-\lambda_{814}/(1+z)}{\lambda_{814}-\lambda_{105}}R_{e,105}\right),
\end{equation}
where $\lambda_r$, $\lambda_{814}$, and $\lambda_{105}$ are the effective wavelengths of the SDSS $r$ band, HST F814W and F105W bands, respectively. We calibrated the FP, as defined in Eq. (\ref{fp}), using $R_e$ measured in kiloparsecs, as described in Eq. (\ref{re}), whilst $\mathrm{SB}_e$ is measured in $\rm mag \,arcsec^{-2}$ as
\begin{equation}
    \mathrm{SB}_e = m_{814} +  K_{814,r} (z) + 2.5 \log(2\pi R_e^2) - 10 \log(1+z),
\end{equation}
where $K_{814,r} (z)$ is defined as in Eq. (\ref{kcorr}), the corrected $R_e$ is in arcsec, and the last term takes into account cosmological dimming.

We selected the ETGs used to calibrate the FP from their red colour and early-type morphology. Following \citet{tortorelli18,tortorelli23}, we calibrated a colour-magnitude relation between the total magnitude in the F814W band and the $\mathrm{F606W}-\mathrm{F814W}$ colour, for all galaxies included in the photometrically completed cluster member catalogue presented in Sect. \ref{sec:cmcat}:
\begin{equation}\label{eqcm}
    m_{606}-m_{814} = a + b \, m_{814}.
\end{equation}
We selected red galaxies as those redder than the best-fit CM relation minus $1\sigma$, with the exception of A2744, where we included galaxies redder than the best-fit CM relation minus $2\sigma$, as the scatter provided by the fitting procedure is significantly smaller than all other clusters and appears underestimated. The best-fit parameters of the colour-magnitude relation for the four clusters are presented in Table \ref{cm}, and are compatible with those found by \cite{tortorelli18} at $1\sigma$ level for AS1063 and at $2\sigma$ level for M1149. We also followed the criterion adopted by \citet{tortorelli18} to identify ETGs from their morphology, requiring a S{\'e}rsic index $n>2.5$. This left us with a sample of 49, 25, 34, and 25 red early-type members for A2744, AS1063, M0416, and M1149, respectively, that can be used as a basis for the calibration of the FP.

\begin{table}
\caption{\label{cm}Parameters of the colour--magnitude relation for the four clusters.}
\centering
\begin{tabular}{lcc}
\hline\hline
Cluster&$a$&$b$\\
\hline
A2744           & $1.61\pm0.06$     & $-0.037\pm0.003$ \\
AS1063           & $1.86\pm0.12$   & $-0.046\pm0.006$ \\
M0416 & $1.69\pm0.11$     & $-0.037\pm0.005$ \\
M1149        & $2.00\pm0.23$      & $-0.036\pm0.011$ \\
\hline
\end{tabular}
\tablefoot{We report, for each cluster, the values of the two colour-magnitude parameters, defined as in Eq. (\ref{eqcm}), with the uncertainty determined by \texttt{ltsfit}.}
\end{table}

\begin{figure}
   \centering
   \includegraphics[width=\hsize]{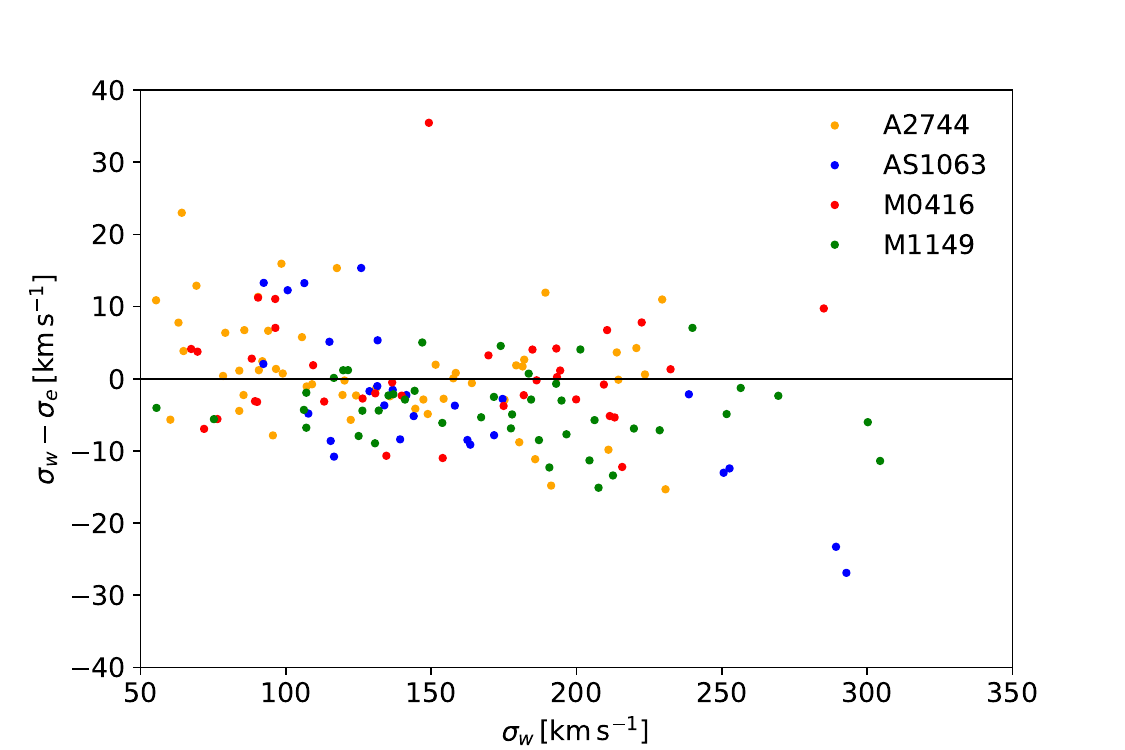}
      \caption{Velocity dispersion values of the cluster members from weighted spectra ($\sigma_w$) compared with the velocity dispersion within the effective radius ($\sigma_e$). The weighted spectrum was extracted within an aperture of $1.5''$ from the galaxy centre of light from the MUSE cubes weighted with the HST F814W surface brightness.}
         \label{compap}
   \end{figure}

\begin{figure*}[t!]
  \includegraphics[width = 0.5\columnwidth]{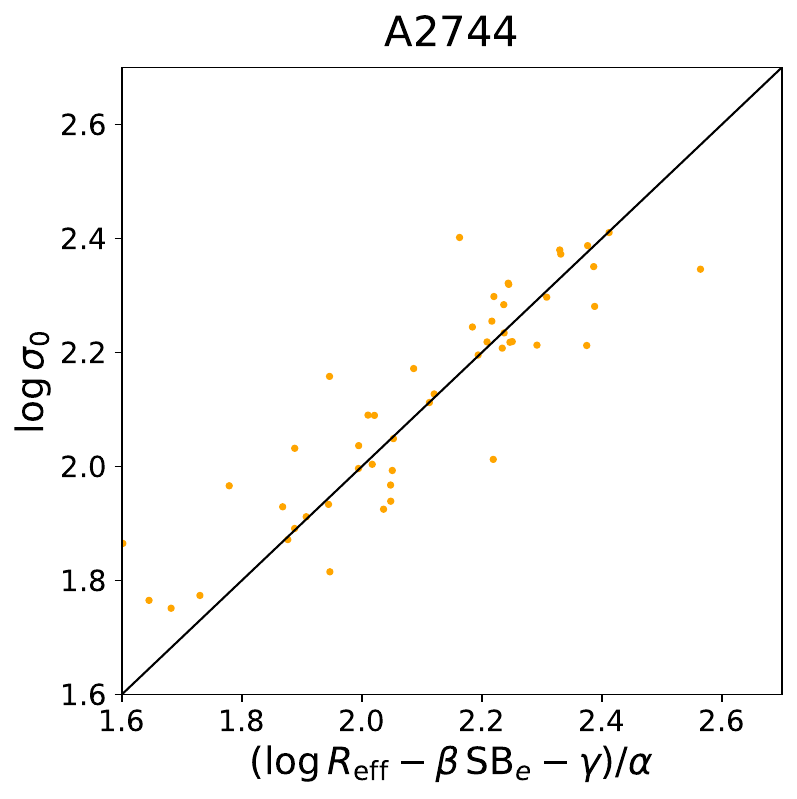}
  \includegraphics[width = 0.5\columnwidth]{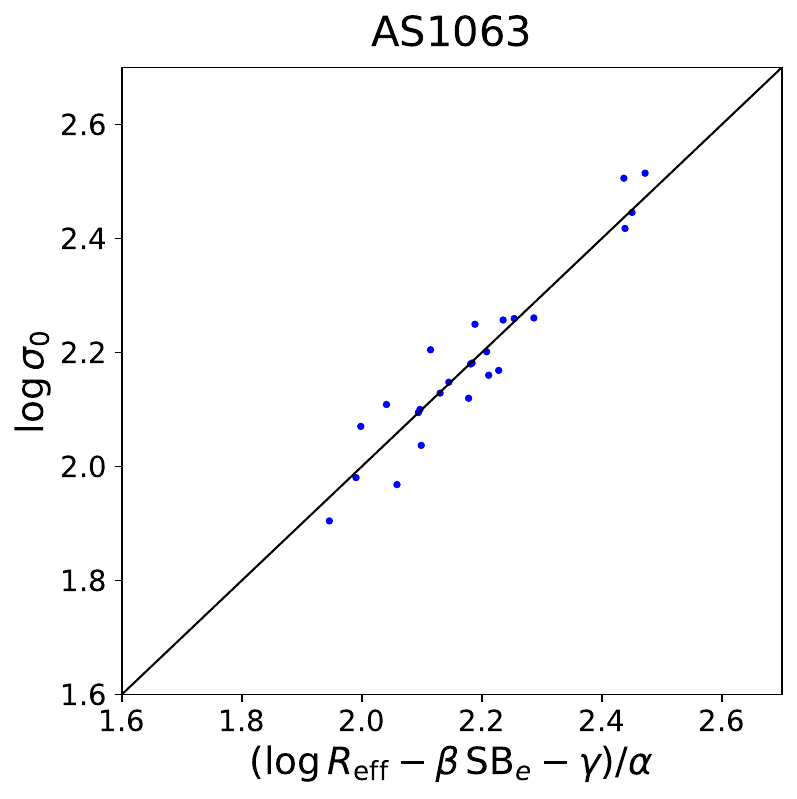}
  \includegraphics[width = 0.5\columnwidth]{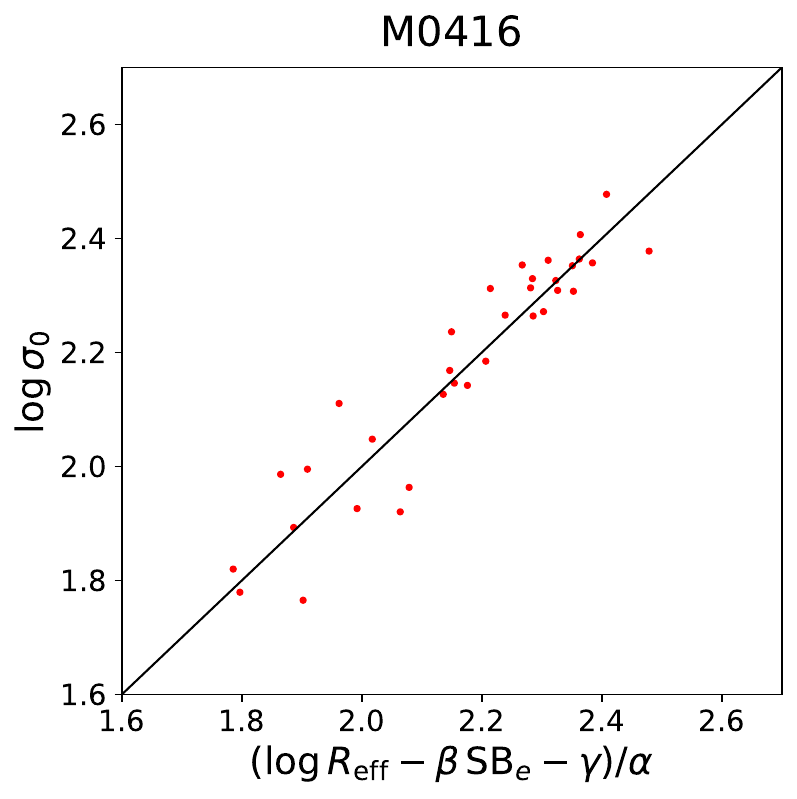}
  \includegraphics[width = 0.5\columnwidth]{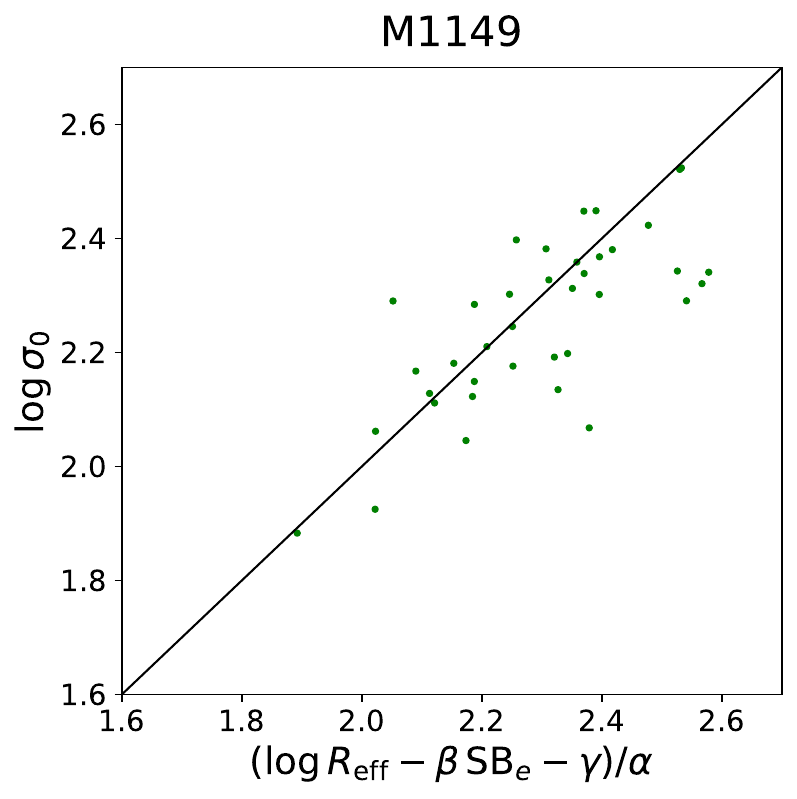}
  \caption{Distribution around the FP relations of the galaxies used for the calibration for the four clusters. The FP relations are shown on the plot diagonal. The values of $\alpha$, $\beta$, and $\gamma$ for the four clusters are reported in Table \ref{tabfp}.}
  \label{fig:fps}
\end{figure*}

As for the velocity dispersion, we need to correct the measured values from the weighted spectra, which we indicate as $\sigma_w$ to the corresponding central value $\sigma_0$ adopted to calibrate the FP. We adopted a power-law aperture correction \citep{jorgensen96}
\begin{equation}\label{apcorr}
    \sigma_\mathrm{ap}(R)=\sigma_\mathrm{ap}(R')\left(R/R'\right)^{\alpha_\mathrm{circ}},
\end{equation}
where the exponent values $\alpha_\mathrm{circ}$ were determined following \citet{zhu23}, and depend on the $r$-band absolute magnitude, $g - i$ colour, and S{\'e}rsic index of the member. The first step required is understanding how to interpret $\sigma_w$ in terms of a physical aperture $\hat{R}$ such that $\sigma_w=\sigma_\mathrm{ap}(\hat{R})$. As is clear from Fig. \ref{compap}, we find that, on average, $\sigma_w$ is a good approximation of the velocity dispersion within the effective radius $\sigma_e=\sigma_\mathrm{ap}(R_e)$, where $\sigma_e$ has been measured correcting the measured $\sigma_\mathrm{ap}(0.8'')$ through Eq. (\ref{apcorr}). Specifically, $\sigma_w-\sigma_e$ is, on average, for the four clusters, $-1.1 \, \mathrm{km \, s^{-1}}$ with a standard deviation of $7.9 \, \mathrm{km \, s^{-1}}$. The weighting technique for spectral light adopted to increase $S/N$ gives more weight to the central pixels, and the resulting weighted combination of their spectra is naturally representative of the region within the effective radius. This represents a second advantage of the procedure, as this region is usually too small to be probed directly from the MUSE cube: $R_e$ typically encircles just a few MUSE pixels and the resulting spectral $S/N$ is very small. In conclusion, we assume $\sigma_e=\sigma_w$ for all members and use Eq. (\ref{apcorr}) to derive
\begin{equation}\label{apcorrsig0}
    \sigma_0=\sigma_\mathrm{ap}(R_e/8)=\sigma_e \times \left(1/8\right)^{\alpha_\mathrm{circ}} = \sigma_w \times \left(1/8\right)^{\alpha_\mathrm{circ}},
\end{equation}
measured in $\mathrm{km \, s^{-1}}$.
Due to the large number of members of the cluster with a measurement of $\sigma_0$, $R_e$, and $\mathrm{SB}_e$, we can calibrate the FP for the four clusters with an optimisation based on the code \texttt{ltsfit} \citep{cappellari13}. In Table \ref{tabfp}, we report the best-fit values of $\alpha$, $\beta$, and $\gamma$, as defined in Eq. (\ref{fp}), while in Fig. \ref{fig:fps} we show the best-fit FP relations and the scatter around them of the members used for their calibration, for the four clusters. We note that, after the corrections performed on the magnitudes and effective radii of the members to obtain comparable values referred to a rest-frame $r$ band, the values of $\alpha$ and $\beta$ are compatible across the four clusters, with the exception of a lower $\beta$ value for M1149. We find larger uncertainties on the FP parameters for AS1063 and M1149, for which our sample is less inclusive of fainter galaxies, due to the shallower MUSE data. Our $\alpha$ and $\beta$ values are also compatible with the FP relation calibrated in the same band ($r$) for large samples of ETGs observed by the SDSS and the Dark Energy Spectroscopic Instrument (DESI) Peculiar Velocity Survey. Specifically, SDSS and DESI report $\alpha$ values of 1.274 and 1.177, and $\beta$ values of $-0.841$ and $-0.793$, for a sample of 34059 and 3110 ETGs, respectively \citep[see][]{howlett22,said25}, and with the FP relation found by \citet{daddona25} in the HST F814W band for a strong lensing cluster at a similar redshift (PLCK G287.0$+$32.9, $z = 0.3833$).

\begin{figure}
   \centering
   \includegraphics[width=\hsize]{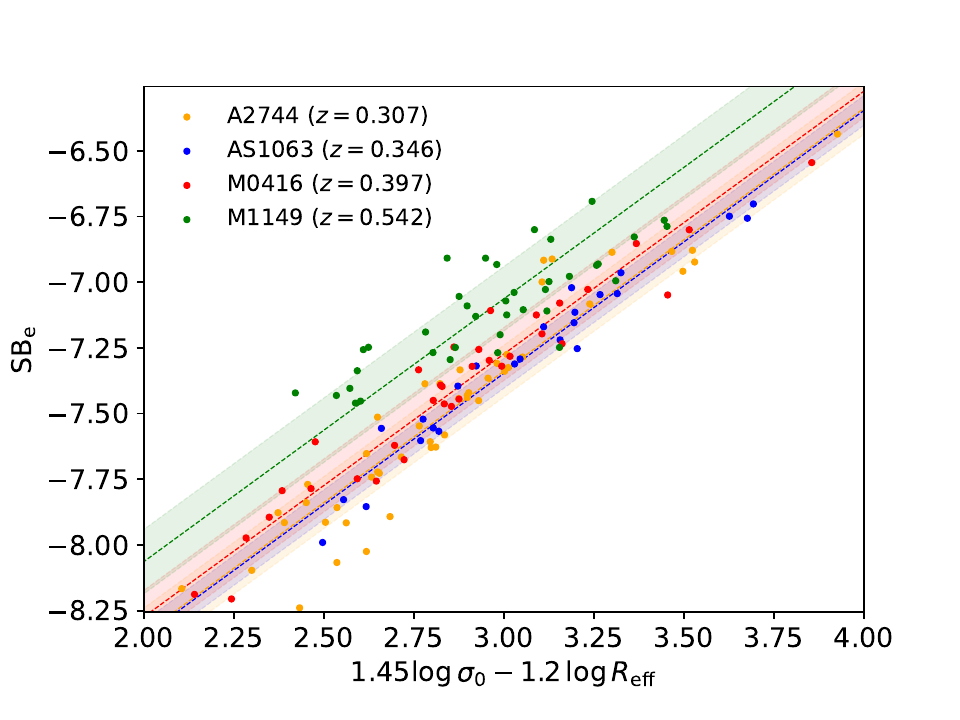}
      \caption{Evolution of the zero-point of the FP relation. We show the FP relations fit for the four clusters assuming $\alpha = 1.20$ and $\beta = -0.83$, and the $1\sigma$ scatter of the galaxies used for its calibration about the plane.}
         \label{fig:vds}
   \end{figure}

The relation between the zero-point of the FP and redshift can be interpreted as a tracer of the evolution of the  mass-to-light ratio ($M/L$) of galaxies, under the assumption that ellipticals are a passively evolving homologous population, so that the $M/L$ evolution with respect to $z=0$ is linked to the variation of $\gamma$ with respect to a FP at $z=0$ by $\Delta \log M/L \, (z) = \Delta \gamma (z)$ \citep[see][]{faber87,vandokkum03,holden05,vandokkum07,vandesande14}. Similarly to \cite{vandesande14} we fitted the FP relations assuming $\alpha = 1.20$ and $\beta = -0.83$, as shown in Fig. \ref{fig:vds}. We find $\Delta \gamma=0.28$ between $z=0.307$ and $z=0.542$, which is larger than the amount predicted by the relation found by \cite{vandesande14}, who reported $\Delta \log M/L \, (z)=(-0.49\pm0.03)z$. Whilst we note that \cite{tortorelli18} also found that the zero-point evolution for the Kormendy relation between AS1063 and M1149 is larger than expected when considering only passive luminosity evolution, the shift between planes is of the order of the $1\sigma$ scatter of the galaxies used for their calibration, limiting the robustness of any conclusion. The study of the evolution of the FP is beyond the scope of this work, due to the limited redshift interval considered, and should be addressed with a larger sample, spanning a wider $z$ range.

\begin{table}
\caption{\label{tabfp}Parameter values of the FP for the four clusters.}
\centering
\begin{tabular}{lcccc}
\hline\hline
Cluster&$\alpha$&$\beta$&$\gamma$&$N$\\
\hline
A2744           & $1.21\pm0.11$     & $-0.87\pm0.05$ & $-8.9\pm0.6$ & 49 \\
AS1063           & $1.03\pm0.30$   & $-0.75\pm0.15$ & $-7.7\pm1.6$ & 25 \\
M0416 & $1.27\pm0.09$     & $-0.85\pm0.04$ & $-8.8\pm0.4$ & 34\\
M1149        & $1.21\pm0.37$      & $-1.19\pm0.31$ &$-11.0\pm2.8$ & 25\\
\hline
\end{tabular}
\tablefoot{We report, for each cluster, the values of the three FP parameters, defined as in Eq. (\ref{fp}), with the uncertainty determined by \texttt{ltsfit}, and the number of galaxies used for the FP calibration.}
\end{table}

\section{The velocity dispersion function of red cluster member galaxies}\label{s5}

The completeness of the velocity dispersion catalogue presented in Sect. \ref{s3} is limited by the smaller MUSE FoV, which only covers the central regions of the cluster core, as well as by the exclusion of members in the region covered by MUSE, but whose spectrum is contaminated by a nearby object. In this section, we take advantage of the FP relations calibrated in Sect. \ref{s4}, and of the larger sets of cluster members described in Sect. \ref{sec:cmcat}, obtained with a combination of spectroscopic and photometric data, to extend the velocity dispersion catalogue to the red members falling within the HFF footprint, including the area not covered by MUSE. As detailed in Sect. \ref{sec:cmcat}, our sample of cluster members is complete across the intersection of all HFF bands, marked in cyan in Fig. \ref{fig:observations}. Whilst the FP strictly only applies to morphologically selected ETGs, we found that including all red galaxies in the FP calibration leads to compatible parameter values for all four clusters, albeit with a larger scatter than with a stricter morphological selection. This is analogous to what found by \cite{tortorelli18} for the Kormendy relation. This suggests that the FP describes in good approximation all red cluster galaxies, and as such we used the calibrated FP relation for each cluster to infer the central velocity dispersion $\sigma_0$ for all the red cluster members falling within the HFF images and use the scatter about the plane to estimate the uncertainty on the recovered values. Similarly to what found by \cite{tortorelli18}, our cluster member sample selected by colour is made up almost exclusively by passively evolving galaxies: fewer than $3\%$ of the red cluster members with a high $S/N$ MUSE spectra ($S/N>10$) shows clear emission lines.

\subsection{Calibration of the velocity dispersion function}\label{cal}
We selected red members with the same criteria described in Sect. \ref{s4}, using the colour-magnitude relations for the four clusters presented in Eq. (\ref{eqcm}) and Fig. \ref{cm}, for a total of 194, 169, 172, and 188 red members for A2744, AS1063, M0416, and M1149, respectively. For these objects we derived $\log \sigma_0$ from $\log R_e$ and $\mathrm{SB}_e$ through Eq. (\ref{fp}). The measured values of $R_e$ and $\mathrm{SB}_e$ are converted into effective values in the rest-frame $r$ band using Eqs. (\ref{kcorr}) and (\ref{re}). As detailed, our sample of cluster members is highly complete up to $m_{814}=25$, so our VDF may suffer from biases due to incompleteness in the bins where members with $m_{814}>25$ fall. We find that setting a threshold $\log \sigma_0 \, [\mathrm{km\,s^{-1}}]>1.5$ ensures a sample completely below $m_{814}=25$ -- even for M1149, the highest-redshift cluster in our sample. This limit is also very close to the minimum measured velocity dispersion value of our catalogue, thus avoiding extrapolating the FP relations beyond the velocity dispersion range used for their calibration.

Having estimated $\sigma_0$ for the red members, we focus on the VDF for each cluster. As detailed, we find an increased scatter about the FP for the full population of red cluster galaxies compared to the sample selected to calibrate it. We used this scatter to evaluate the uncertainty on the $\sigma_0$ values inferred through the FP and how it propagates on the VDF. For each cluster, we took the $1\sigma$ scatter of all red galaxies about the FP in terms of $\log \sigma_0$ and used it to extract 100 values of $\sigma_0$ from a Gaussian distribution for all red member galaxies. After choosing some bins in $\log \sigma_0$, this provided us, for each cluster, with 100 realisations of the number counts of red members in a given velocity dispersion bin. Finally, with the median, and with the 16th and 84th percentiles of the distribution of number counts, we estimated the best value for the VDF in each bin and its $1\sigma$ uncertainty.

\begin{figure}
   \centering
   \includegraphics[width=\hsize]{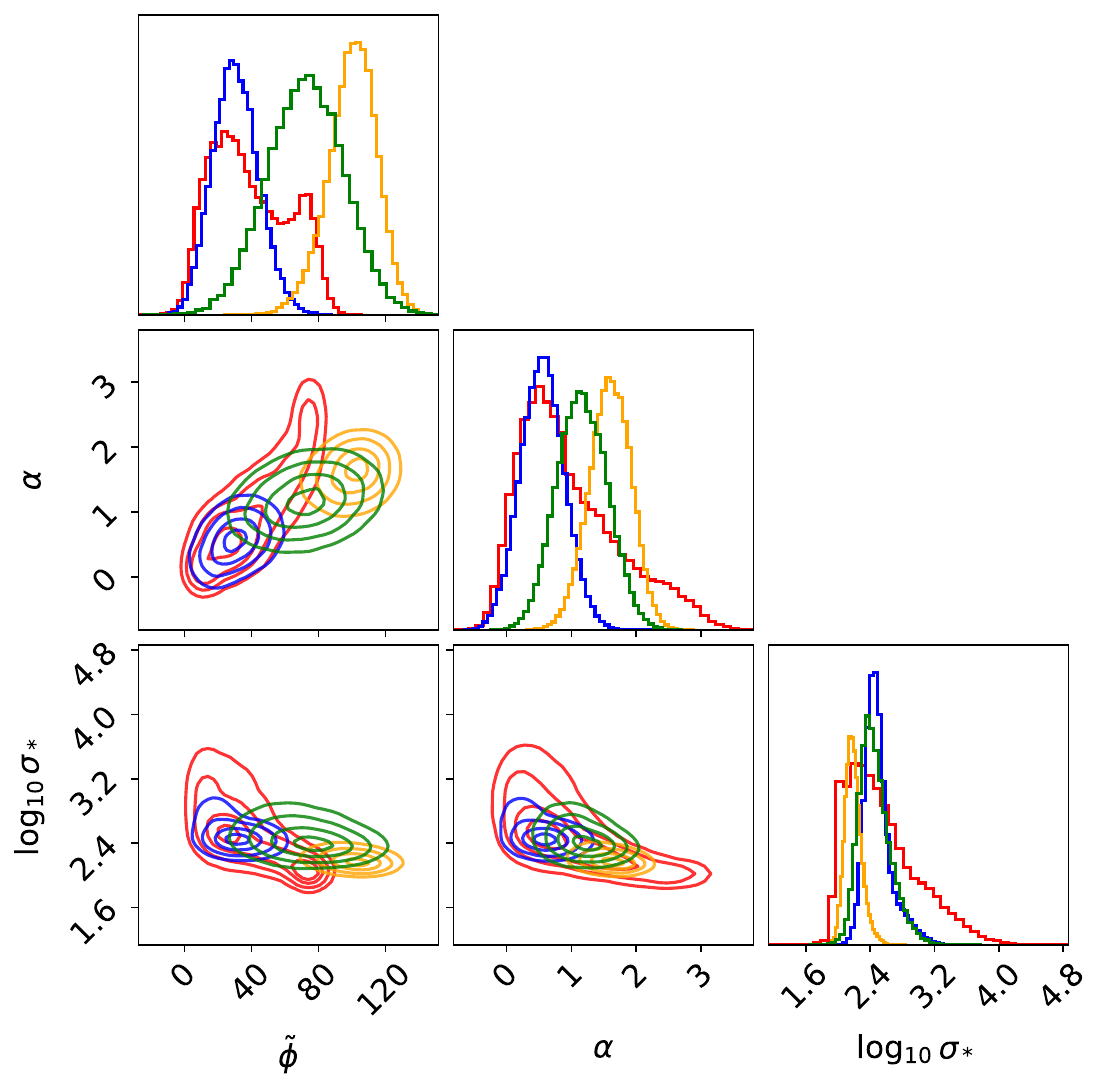}
      \caption{Posterior probability distribution for the parameters of the Schechter function fit of the VDF for the four clusters, as sampled by MCMC chain. The parameters are defined in Eq. (\ref{sfunc}); the colour scheme identifies the four clusters following Fig. \ref{sfuncplot}.}
         \label{postdist}
   \end{figure}

\begin{figure}
   \centering
   \includegraphics[width=\hsize]{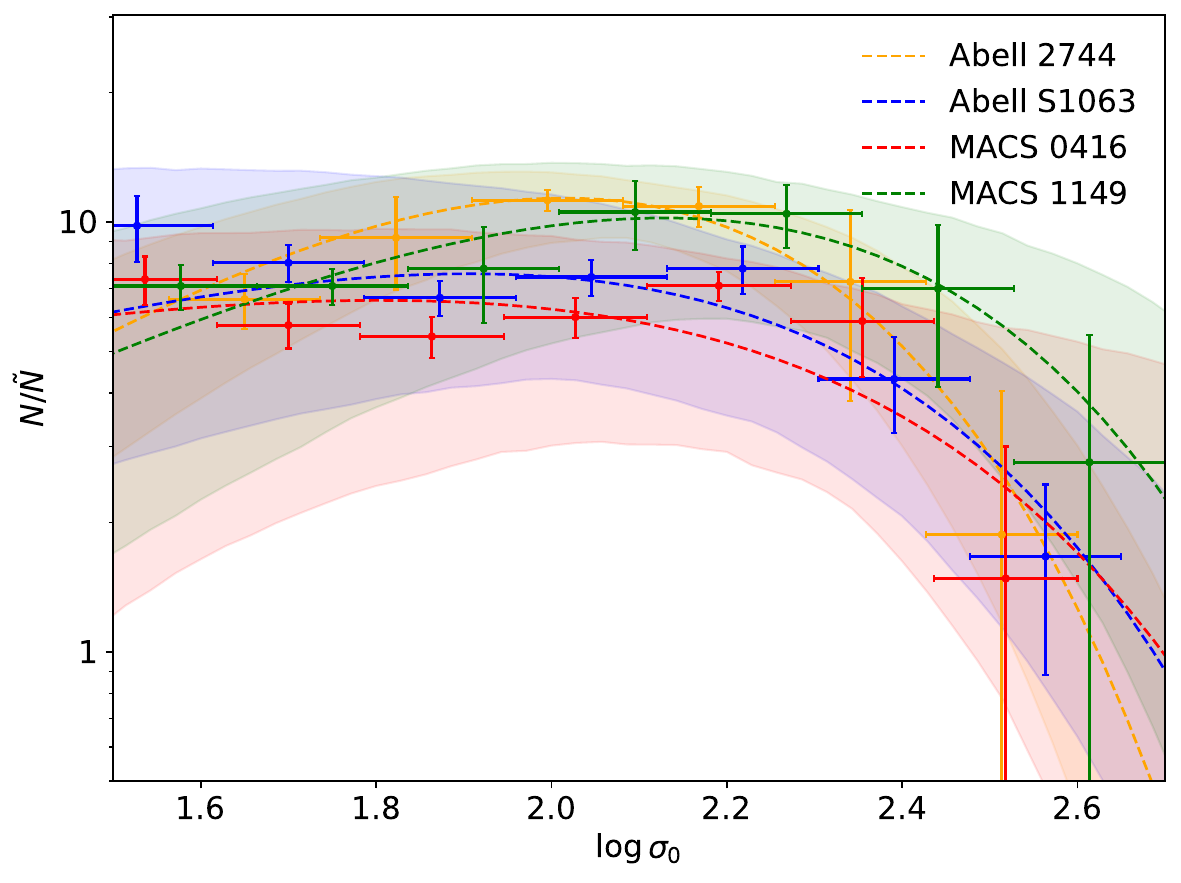}
      \caption{Velocity dispersion function for the four clusters. We show the data bins and the Schechter function fit with their uncertainties, as defined in Eq. (\ref{sfunc}).}
         \label{sfuncplot}
\end{figure}

Following \citet{sohn20}, we describe the VDF of red cluster galaxies with a Schechter function \citep[][]{schechter76}
\begin{equation}\label{sfunc}
    \phi(V)=\phi^*10^{\alpha(V-\log\sigma^*)} \exp\left(-10^{(V-\log\sigma^*)}\right),
\end{equation}
where $\phi(V)$ is the number of galaxies falling in a bin centred in $V=\log \sigma_0$, $\alpha$ is the slope of the power-law regime describing the faint end of the function (for $V<<\log\sigma^*$) and $\log\sigma^*$ marks the transition to an exponentially decreasing regime for the bright end. Similarly to \citet{sohn20}, we find that we cannot simultaneously constrain all parameters of a modified Schechter function parametrisation (Eq. \ref{modsch}) without parametric degeneracy. Compared with Eq. (\ref{modsch}), we are thus setting $\beta=1$, and evaluating the function on finite $V$ bins. The normalisation parameter $\phi^*$ depends on the richness of the cluster, as well as on the ratio between the cluster region covered by all HFF bands, where our cluster member catalogue is complete, and the area within the cluster virial radius $R_{200}$. Focussing on AS1063, M0416, and M1149, \citet{angora20} found an approximate self-similarity between the radial distribution of the cluster members when the number counts for a radial bin at radius $R$ are normalised by the total number of members within $0.15 \, R_{200}$, $N_0$, and by the ratio between the areas enclosed within the radius $R$ and within $R_{200}$. In other words, the number of members in a given area within the cluster core is approximately the same when normalised by the fraction of the virial area covered and the richness of the cluster core. We adopted this approach to allow a comparison between the velocity functions of the four clusters, normalising the VDF in each cluster by
\begin{equation}\label{norm}
    \tilde{N}=N_0 \frac{A_\mathrm{HFF}}{\pi R_{200}^2},
\end{equation}
where $A_\mathrm{HFF}$ is the area covered by the intersection of all HFF bands. For clarity, we use $\tilde{\phi}$ to refer to the value of $\phi^*$ for the normalised Schechter function. We used the values of $R_{200}$ by \citet{medezinski16} for A2744 and those by \citet{umetsu18} for the remaining three clusters, following \citet{angora20}. Given that the region covered by HFF does not extend fully to $0.15 \, R_{200}$ for all clusters, we chose the number of red members within $0.1 \, R_{200}$ as a proxy of richness ($N_0$). We report the values of $\tilde{N}$ in Table \ref{sfpar}. We remark that this self-similarity only holds approximately and is potentially influenced by the morphology and projected structure of the clusters. We fitted Eq. (\ref{sfunc}) with an affine-invariant Markov chain Monte Carlo \citep[MCMC;][]{goodman10,foreman13} algorithm. We explored the posterior probability distribution for the values of $\tilde{\phi}$, $\alpha$, and $\log\sigma^*$ as defined in Eq. (\ref{sfunc}) and in the text, expressed as a product between a likelihood
\begin{multline}\label{lik}
    \ln \left( \mathcal{L}(\phi(V)| V,\delta\phi(V),\tilde{\phi},\alpha,\log\sigma^*) \right) = - \frac{1}{2} \sum_{i=1}^{N_\mathrm{bins}}  \frac{\left(\phi(V_i)-\hat{\phi}(V_i)\right)^2}{\left(\delta\phi(V_i)\right)^2} ,
\end{multline}
and flat priors for all the free parameters, chosen to limit the parameter space. In Eq. (\ref{lik}), $\phi(V_i)$ and $\delta\phi(V_i)$ are the median and $1\sigma$ uncertainty of the distribution of number counts in a given bin $V_i$ in $\log \sigma_0$ space, normalised by the factor $\tilde{N}$, introduced in Eq. (\ref{norm}), whilst $\hat{\phi}(V_i)$ are the model-predicted values in the same bin for a Schechter function model with parameters $\tilde{\phi}$, $\alpha$, and $\log\sigma^*$. We notice that some bins in the high-$V$ regime only contains a few objects, and that this causes the recovered Schechter function parameters to be sensitive to the binning choice. To take this source of uncertainty into account, we took nine realisation of the bins $\left\{ V_i \right\}_{i=1}^{N_\mathrm{bins}}$, shifted by $\left\{-0.5;-0.375;-0.25;-0.125;0;0.125;0.25;0.375;0.5 \right\} \times \Delta V$, where $\Delta V$ is the bin width. Then we repeated the MCMC optimisation for each bin shift. The final posterior probability distributions for the parameters are the sum of the nine distributions obtained for each of the realisation. We took the median, 16th, and 84th percentiles of the sampled posterior probability distributions to estimate the best-fit values of the optimised parameters for all clusters and their uncertainties, reported in Table \ref{sfpar}. The posterior probability distribution of the three parameters for the four clusters, as sampled by MCMC chains, is instead shown in Fig. \ref{postdist}. The values of $\alpha$ and $\log\sigma^*$ are always compatible at $1\sigma$ or $2\sigma$ level, and Fig. \ref{postdist} shows a clear overlap and alignment of the corner-plots between them, suggesting a similar correlation for the two parameters across the four clusters. In general, the results favour a positive $\alpha$ slope in the range $0.55-1.60$ and $\log\sigma^*\, [\mathrm{km\,s^{-1}}]$ between $2.18$ and $2.47$, corresponding to $\sigma^*$ values of $151-295 \, \mathrm{km \, s^{-1}}$. On the other hand, the distributions of $\tilde{\phi}$ are brought closer by the normalisation adopted (see Eq. \ref{norm}), but the scatter between the clusters remains more significant, which points towards intrinsic differences in the cluster richness. We extracted correlated sets of parameters from the posterior probability distributions shown in Fig. \ref{postdist} to plot the VDFs with their propagated uncertainty in Fig. \ref{sfuncplot}. The bin shifts in the optimisation procedure are reflected in the bin width in Fig. \ref{sfuncplot}. We note that the posterior probability distributions for M0416 appear broader and occasionally double-peaked. This behaviour arises because, for some bin configurations, the highest-$V$ bin becomes empty and is consequently excluded from the fit -- giving rise to a second peak in the parameter space for the posterior probability distribution. As such, uncertainty on the parameters for M0416 may be overestimated. We also note that M0416 is a distinctly bi-modal cluster (see Fig. \ref{fig:observations}), with two sub-clusters clearly emerging, which might determine a different evolution history for its member galaxies, affecting the VDF.

\begin{table}
\caption{\label{sfpar}Parameters of the best Schechter function fit of the velocity dispersion function for the red members of the four clusters.}
\centering
\begin{tabular}{lcccc}
\hline\hline
Cluster&$\tilde{\phi}$&$\alpha$&$\log\sigma^*\, [\mathrm{km\,s^{-1}}]$&$\tilde{N}$\\
\hline
\vspace{1 mm}
A2744           & $101^{+13}_{-15}$     & $1.60^{+0.32}_{-0.33}$ & $2.18^{+0.11}_{-0.09}$ & 2.14  \\
\vspace{1 mm}
AS1063           & $30^{+14}_{-13}$     & $0.55^{+0.34}_{-0.34}$ & $2.47^{+0.18}_{-0.12}$ & 2.55\\
\vspace{1 mm}
M0416 & $36^{+32}_{-20}$     & $0.80^{+1.01}_{-0.56}$ & $2.46^{+0.57}_{-0.34}$ & 2.95 \\
\vspace{1 mm}
M1149        & $73^{+22}_{-23}$     & $1.16^{+0.40}_{-0.39}$ & $2.43^{+0.22}_{-0.16}$ &2.54 \\
\hline
\end{tabular}
\tablefoot{We report, for each cluster, the values of the three Schechter function parameters, defined as in Eq. (\ref{sfunc}), with the uncertainty determined by the MCMC optimisation, and the normalisation $\tilde{N}$ used for the velocity dispersion bins.}
\end{table}

\subsection{Comparison between the calibrated functions and with other works}

\begin{figure}[t!]
  \centering
  \includegraphics[width=\columnwidth]{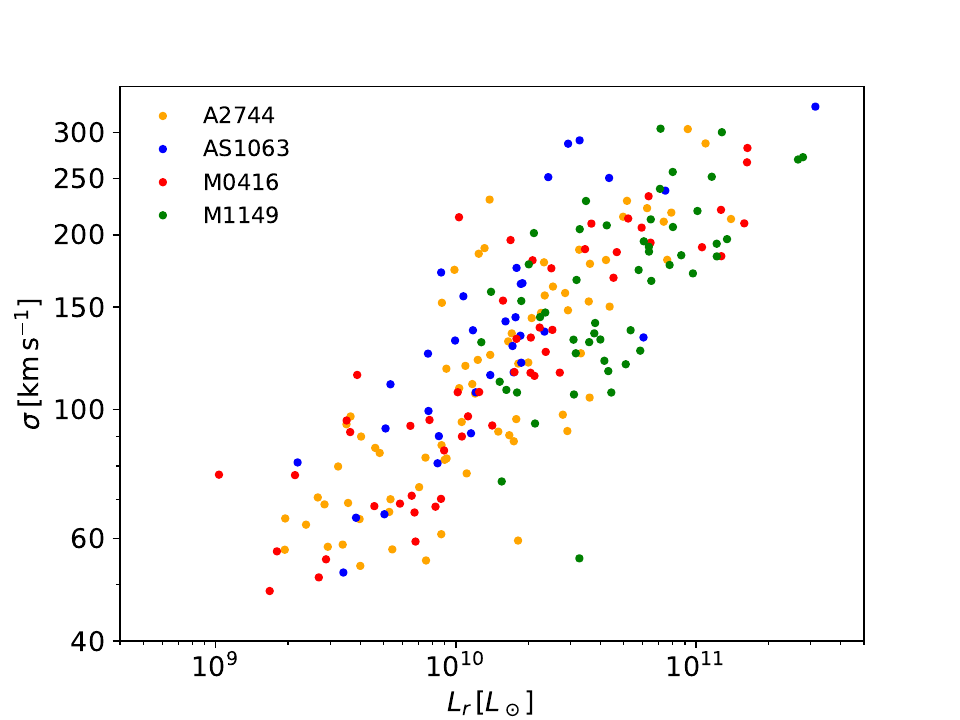}
  \caption{Measured velocity dispersion as a function of rest-frame SDSS $r$-band luminosity for our work.}
  \label{fig:siglum}
\end{figure}

\begin{figure}[t!]
  \centering
  \includegraphics[width=\columnwidth]{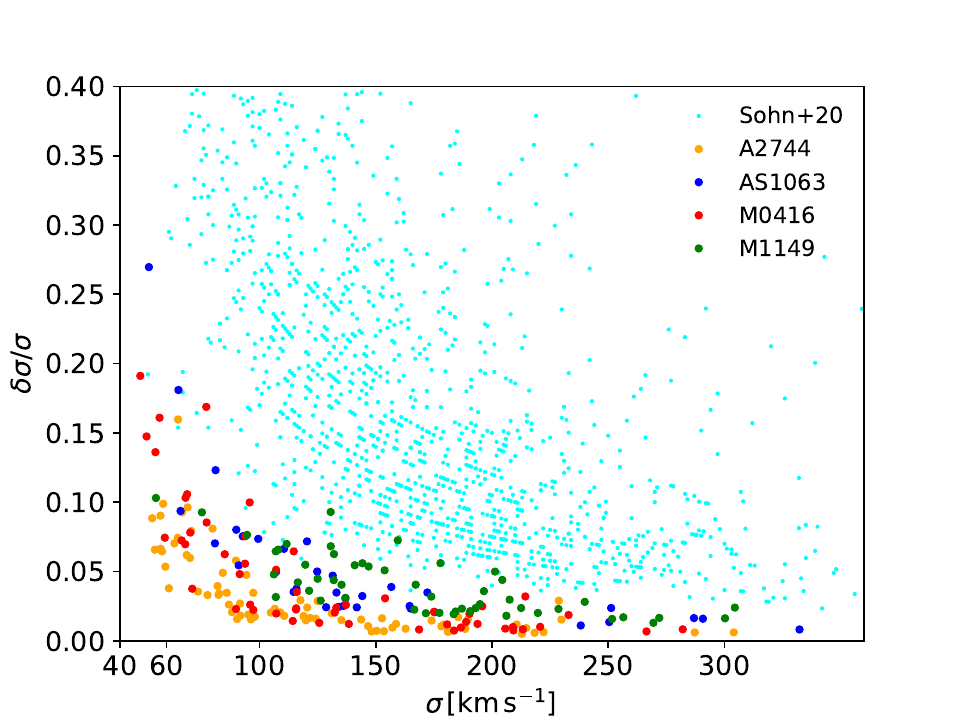}
  \caption{Relative uncertainty on the measurement of the velocity dispersion. Comparison between the $\sigma_w$ values presented in this work and the Hectospec observations of nine strong lensing clusters \citep[][in cyan]{sohn20}.}
  \label{fig:sohn}
\end{figure}

As pointed out in the previous sub-section, the best-fit values of $\alpha$ and $\log\sigma^*$ for a Schechter function-fit of the normalised VDFs are compatible across the four clusters. However, some differences between the four VDFs emerge when looking at the higher-velocity-dispersion regime. Whilst for $\log\sigma_0\, [\mathrm{km\,s^{-1}}]<2.2$ data bins for the four functions are compatible and very close to one another, M1149 shows some excess for $\log\sigma_0\, [\mathrm{km\,s^{-1}}]>2.2$. Whilst any conclusion is limited by the small cluster sample, it is interesting to note how M1149 has a significantly higher virial mass (see Table \ref{sfpar}) and redshift compared to the other three. This may be a trace of major dry dissipationless mergers \citep[see e.g.][]{khochfar03,bolyan06,delucia07,naab09} reducing the number of high-velocity-dispersion galaxies in lower-$z$ clusters. This hypothesis is also suggested by the higher BCG velocity dispersion values we find for the lower-$z$ clusters A2744 and AS1063 (see Tables \ref{tab:table2744}, \ref{tab:table2248}, \ref{tab:table0416}, \ref{tab:table1149}), which can also be a hint of major dry mergers affecting the BCG structure \citep[][]{cox06}.

The integral-field nature of MUSE allows us to simultaneously measure the spectra of all cluster members in the core with very few pointings, efficiently allowing deep exposures. As shown in Fig. \ref{fig:siglum}, we can reliably extend our catalogue to values well below $100 \mathrm{\,km \, s^{-1}}$, corresponding to a total rest-frame SDSS $r$-band luminosity $\gtrsim 10^9 \, L_\odot$, probing galaxy mass and luminosity regimes unexplored at these redshifts. Figure \ref{fig:siglum} also allows us to probe the slope of power-law scaling relations between $\sigma$ and the total luminosity, akin to those adopted in SL, albeit in a different band: we find that a fit of $\sigma\propto L_r^\alpha$ gives $\alpha$ values in the range $0.30-0.38$ for the four clusters, compatible with those found by cluster-scale SL models \citep[see][]{granata25}. Such low-luminosity targets were inaccessible to the velocity dispersion survey from Hectospec spectra used to calibrate the VDF in \citet{sohn20} for a set of nine strong lensing clusters in the redshift range $0.18-0.29$ (see Fig. \ref{fig:sohn}). Looking at the VDF parameters, \citet{sohn20} found higher values of $\alpha$, between $1.75$ and $4.40$, and lower values of $\log\sigma^*$, in the range $1.46-1.72$, compared to our work. Due to the stricter lower limit $\log\sigma\, [\mathrm{km\,s^{-1}}]>2.05$ on the velocity dispersion values they consider, the recovered parameter values may be influenced by the fact that $\log\sigma^*$ falls outside the velocity dispersion range considered. In our work, on the other hand, the depth of the MUSE observations has allowed us to extend the VDF to a wider velocity dispersion range, accurately sampling its power-law regime. \citet{sohn20} also found that higher-mass galaxy clusters have higher $\sigma$ values for the BCG, but fewer high-$\sigma$ members, which may be a trace of major dry mergers, similarly to what we found for lower $z$. We do not find any such trend with the virial mass values of the clusters ($2.06 \times 10^{15} M_\odot$, $1.98\times 10^{15} M_\odot$, $1.14\times 10^{15} M_\odot$, and $2.89\times 10^{15} M_\odot$ for A2744, AS1063, M0416, and M1149, respectively) but we should remark that any trend emerging from a comparison between the four clusters is significantly limited by the small sample.
   
\section{Summary and results}\label{s6}

We have presented a detailed and comprehensive study of the kinematic properties of the red member galaxies in the cores of four strong lensing clusters at $z = 0.307-0.542$ chosen from the HFF sample, for which exquisite photometry and deep integral-field spectroscopy from MUSE are available. Taking advantage of the MUSE data, we built a kinematic pipeline based on the spectral fitting code \texttt{pPXF} aimed at reliably measuring the velocity dispersion of cluster members falling within the instrument FoV. We tested our pipeline with a library of 16,000 simulated spectra (see Appendix \ref{appa}), with the objective of a more accurate assessment of the uncertainty on our measurements. Using a library of 462 XSL DR3 spectra, selected in \citet{knabel25}, we found that \texttt{pPXF} allows us to measure $\sigma$ with a statistical uncertainty consistently below $5\%$ when the spectral $S/N$ is higher than 10. For higher $S/N$ spectra, even lower uncertainties can be achieved for the brightest galaxies, potentially of the order of $1\%$.

We introduced a spectral light weighting technique that exploits the HFF images to increase the $S/N$ and introduce a proxy for the galaxy light distribution, which acts as an approximate aperture correction to the effective radius of the member galaxy. With the pipeline in place, we built an extensive catalogue of 213 measured velocity dispersion values across the four clusters, which we report in Appendix \ref{appc}. This sample is unprecedented in terms of combined depth and accuracy for cluster galaxies at this redshift. The pipeline and the procedure presented in this work can be applied to gain kinematic priors to calibrate the scaling relations that describe the total mass distribution of cluster members in strong lensing models of galaxy clusters and has been used in several studies for this purpose \citep[][Bianchetti et al. in prep.]{bolamperti24,granata25,ertl25,acebron25,abriola25}.

We took advantage once again of the deep multi-band HFF photometry to extend the catalogue of measured velocity dispersion to all red members, including those whose spectrum is contaminated by a nearby source or has an insufficient $S/N$, and those falling outside of the footprint of the MUSE data cubes but within that of the HFF observations. We measured the structural parameters using \texttt{MORPHOFIT} for an extensive sample of 723 red cluster members within the four clusters, selected either spectroscopically or photometrically using a convolutional neural network. The sample is highly pure and complete down to $m_\mathrm{F814W}=25$. In order to do so, we calibrated the FP relation for all four clusters in the rest-frame $r$ band, using $K$-corrected magnitudes and effective radii obtained from the multi-band structural parameters. The best-fit values of the parameters $\alpha$ and $\beta$ of the FP relations for the four clusters are compatible with each other (see Table \ref{tabfp}), and with FP relations calibrated for samples of ETGs in the same band. The third parameter $\gamma$ shows the expected signs of evolution with redshift. The offset we find across the redshift range considered is larger than expected under the effect of passive luminosity evolution only, similarly to what was found for the Kormendy relation zero-point \citep{tortorelli18}, and could point to hints of size evolution. However, the small sample and the $z$ range considered prevent us from drawing any strong conclusions. 

With the calibrated FP relations and their scatter, we assigned a velocity dispersion to all 723 red cluster members and built a VDF for each cluster, extended down to $\log \sigma_0\, [\mathrm{km\,s^{-1}}]=1.5$. We used the number of members within $R=0.1R_{200}$ and the ratio of the area covered by HFF observations to the virial area to normalise the functions, and we fitted a Schechter function to the relations; we found that they are well described by a positive $\alpha$ slope with values in the range $0.55-1.60$ and $\log\sigma^*\, [\mathrm{km\,s^{-1}}]$ between $2.18$ and $2.47$. The two parameters $\alpha$ and $\log\sigma^*$ are compatible across the four clusters; however, the highest-$z$ cluster of the sample, M1149, shows an excess for $\log\sigma_0\, [\mathrm{km\,s^{-1}}]>2.2$, which might be explained by major dry mergers reducing the number of high-$\sigma_0$ member galaxies at lower redshifts, whilst increasing the velocity dispersion of the BCG. On the other hand, we find significant differences with $\alpha$ and $\log\sigma^*$, as found for nine strong lensing clusters at $z=0.18-0.29$ by \citet{sohn20}, which may depend on limitations descending from their lower bound $\log\sigma_0\, [\mathrm{km\,s^{-1}}]>2.05$. Comparing our results with the local massive clusters Coma and Abell 2029 \citep[][]{sohn17} to probe the redshift evolution of the cluster-core VDF is not straightforward, given the different parametrisation chosen. However, a side-by-side comparison reveals broadly similar slopes for the VDFs in the low-$\sigma$ power-law regime. This suggests that the abundance of intermediate-mass early-type galaxies in cluster cores does not evolve strongly between $z\sim0$ and the redshift range probed by the HFF clusters. Our conclusions on the VDF are, in any case, limited by the small sample, as highlighted by the cluster-to-cluster scatter within our sample, and by the limited coverage of the HFF observations, extended to $R\lesssim0.15R_{200}$, but explore velocity dispersion regimes mostly uncharted in previous studies. 

The deep pan-chromatic observations and MUSE integral-field spectroscopy now available for the cores of massive strong lensing clusters have allowed us to carry out an extremely detailed and comprehensive study of the kinematics of red cluster galaxies, accurately probing the depth of the potential wells of the sub-haloes that host them. Luminosity-function-based studies depend on stellar population properties (age, metallicity, star formation history) and dust attenuation, and calibrating the stellar mass function of cluster galaxies requires assumptions on their mass-to-light ratios and initial mass functions \citep[e.g.][]{bernardi10,monterodorta17}. Furthermore, the luminosities and mass-to-light ratios of cluster galaxies are affected by quenching effects arising from the dense cluster environment \citep[e.g. ram-pressure stripping and starvation, see][]{gunn72,larson80,peng10b}, which do not substantially modify their total gravitational potential, thus making the luminosity and stellar-mass functions environment dependent tracers of total mass. In contrast, the central stellar velocity dispersion of galaxies is inferred directly from the width of their spectral absorption lines, and closely traces the depth of the gravitational potential in their inner regions, providing a much more straightforward tracer of the total mass distribution of cluster galaxies, and thus of the properties of the DM haloes that host them. This allows  a direct comparison with the mass properties of cluster galaxies as inferred by strong lensing models, which are likewise sensitive to the structure of their inner potential wells. This will crucially inform attempts at resolving the persistent excess of galaxy-galaxy strong lensing events in galaxy clusters with respect to cosmological hydrodynamical simulations \citep[][]{meneghetti20,meneghetti22}, a discrepancy that also depends on the detailed mass structure of cluster sub-haloes. Whilst this work focussed on a small sample with homogeneous photometry, extending it to a larger sample, spanning a wider redshift range, will allow us to test the robustness of our conclusions and meaningfully probe emerging evolutionary trends, thus constraining galaxy formation and evolution scenarios in clusters.

\section*{Data availability}

The catalogues of structural parameters presented in Appendix \ref{appendix:cat_description} are available at the CDS via anonymous ftp to cdsarc.cds.unistra.fr or via https://cdsarc.cds.unistra.fr/viz-bin/cat/J/A+A/709/A254.

\begin{acknowledgements}
We thank Bodo Ziegler for the useful feedback and comments that helped to improve the quality of the manuscript. We acknowledge support from the Italian Ministry of University and Research through grant PRIN-MIUR 2020SKSTHZ. PB acknowledges support from the Italian Space Agency (ASI) through contract ``Euclid - Phase E'', INAF Grants ``The Big-Data era of cluster lensing'' and ``Probing Dark Matter and Galaxy Formation in Galaxy Clusters through Strong Gravitational Lensing.''
This work uses the following software packages:
\href{https://github.com/astropy/astropy}{\texttt{Astropy}}
\citep{astropy1, astropy2},
\href{https://github.com/matplotlib/matplotlib}{\texttt{matplotlib}}
\citep{matplotlib},
\href{https://github.com/numpy/numpy}{\texttt{NumPy}}
\citep{numpy1, numpy2},
\href{https://pypi.org/project/ppxf/}{\texttt{pPXF}}
\citep{cappellari04,cappellari23},
\href{https://github.com/torluca/morphofit}{\texttt{MORPHOFIT}}
\citep{tortorellimercurio,tortorelli23},
\href{https://www.python.org/}{\texttt{Python}}
\citep{python},
\href{https://github.com/scipy/scipy}{\texttt{Scipy}}
\citep{scipy}.
\end{acknowledgements}

% WARNING
%-------------------------------------------------------------------
% Please note that we have included the references to the file aa.dem in
% order to compile it, but we ask you to:
%
% - use BibTeX with the regular commands:
\bibliographystyle{aa} % style aa.bst
\bibliography{bibl.bib} % your references Yourfile.bib
%
% - join the .bib files when you upload your source files
%-------------------------------------------------------------------
\clearpage            % flush all earlier floats
\begin{appendix}
\section{Testing the accuracy of the velocity dispersion measurements with simulated spectra}\label{appa}

In this appendix we test the reliability of the kinematic measurements performed with \texttt{pPXF} on a set of simulated MUSE spectra of member galaxies. Here and in the main body of the paper, we refer to the $S/N$  measured as an average over the studied wavelength range of the ratio between the observed spectrum and its residuals with respect to the best fit found by \texttt{pPXF}, computed within $150 \, \mathrm{\AA}$ bins.

\begin{figure}
   \centering
   \includegraphics[width=\hsize]{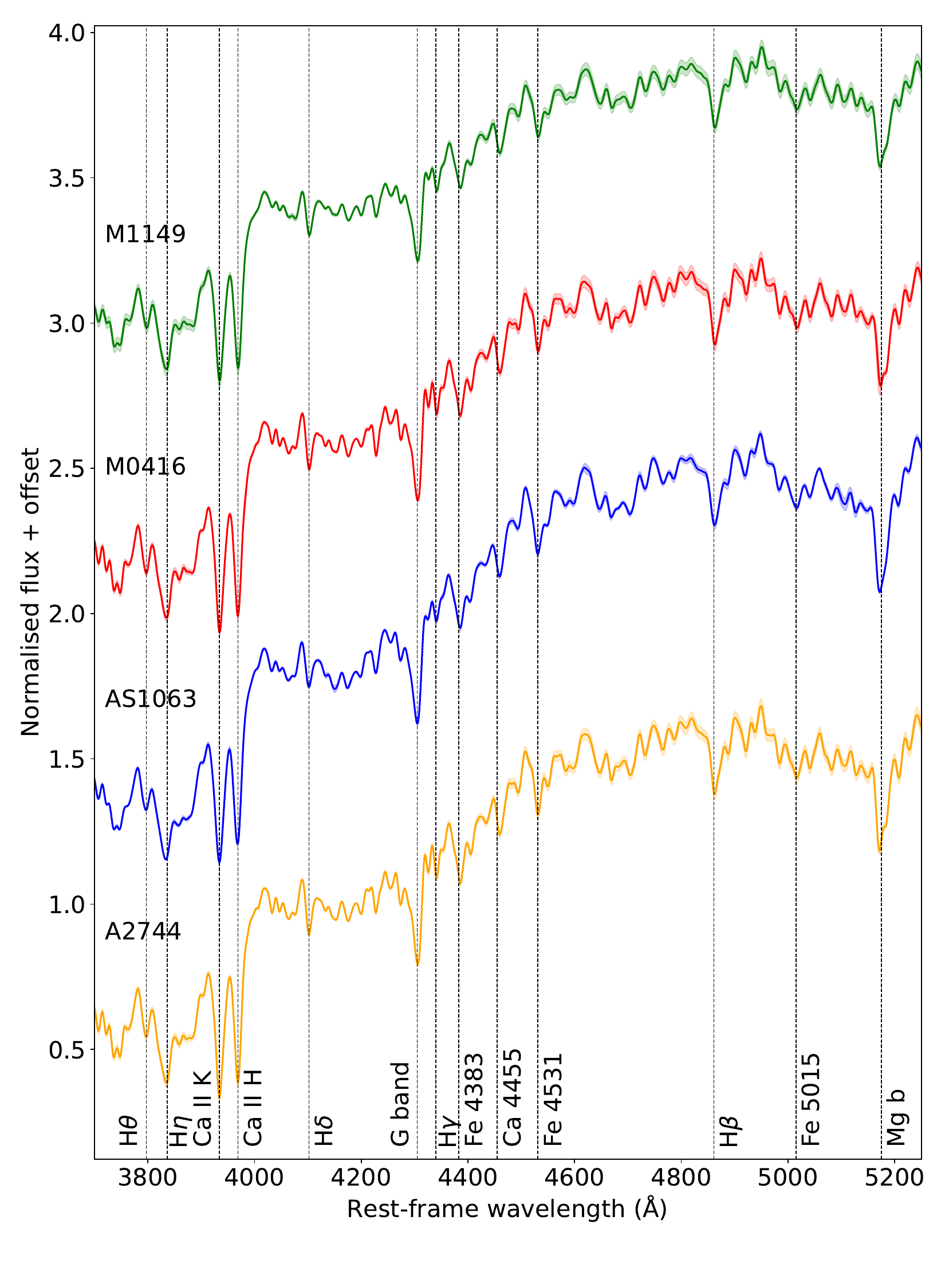}
      \caption{VLT/MUSE mean stacked normalised spectra of 14, 6, 17, and 27 member galaxies of A2744, AS1063, M0416, and M1149, respectively, with $\sigma > 180 ~{\rm km~s^{-1}}$ and $S/N > 15$. The shaded regions are the standard deviation of each spectral pixel. The spectra are smoothed with a Gaussian kernel with a standard deviation of $3.75 ~\mathrm{\AA}$. The red dashed vertical lines locate the absorption features detected in the spectra. For  clearer visualisation, the normalised flux values from AS1063, M0416, and M1149 have been increased by 1, 2, and 3, respectively.}\label{showspectrapp}
\end{figure}

We generated simulated spectra of cluster galaxies using our MUSE observations as empirical templates. Specifically, we adopted the best-fit combinations of X-Shooter stellar templates derived with \texttt{pPXF}-fitting for a sample of 40 high $S/N$ members of M0416 for which robust velocity dispersion measurements are available. Figure \ref{showspectrapp} shows the quality of our spectra and the available absorption features. These spectra serve as reference bare galaxy spectra at $\sigma=0$, with a resolution matching that of the templates ($R \approx 10000$).

To construct the simulated sample, we randomly selected one of 40 bare spectra and re-sampled it onto a wavelength grid ten times finer to avoid interpolation artifacts. We then convolved each oversampled spectrum with a Gaussian LOSVD, parametrised by a velocity $v$ and dispersion $\sigma$, drawn from distributions shown in Table \ref{simpar}. The spectrum was redshifted to $z=0.4$, degraded to the MUSE resolution (FWHM $= 2.6\,\mathrm{\AA}$), and re-sampled to the MUSE pixel scale ($\Delta\lambda = 1.25\,\mathrm{\AA}$). Random Gaussian noise was added to match a specified signal-to-noise ratio. Velocity values were sampled uniformly in the range $v \in [-50, 50]\,\mathrm{km\,s^{-1}}$, while $\sigma$ and S/N were drawn from the intervals listed in Table~\ref{simpar}.

Given that the sample of galaxies selected for our velocity dispersion catalogues does not push below $30 \, \mathrm{km \, s^{-1}}$, we adopted it as lower limit for our simulation sample and for the range on which we tested the reliability of our pipeline. The mock galaxy spectra were fitted in the same rest-frame wavelength range of $[3700-5250] \, \mathrm{\AA}$ and the same \texttt{pPXF} set-up described in Sect. \ref{s3}. As anticipated in the text, we tested different degrees for the additive Legendre polynomials used to correct for the continuum shape, and find the fits to be largely insensitive to the choice \citep[similarly to][]{cappellari17}. We fixed the degree to 12 to avoid possible oscillations of the recovered values for low-degree polynomials \citep[as found by][]{dago23}. Both these fits and those of observed cluster members show that for $S/N<10$ the fit often completely fails, or struggles to recover the observed shape of the absorption features. As such, we  disregarded cluster members with a spectral $S/N<10$ for our velocity dispersion measurements.

\begin{figure*}[t!]
  \includegraphics[width = \columnwidth]{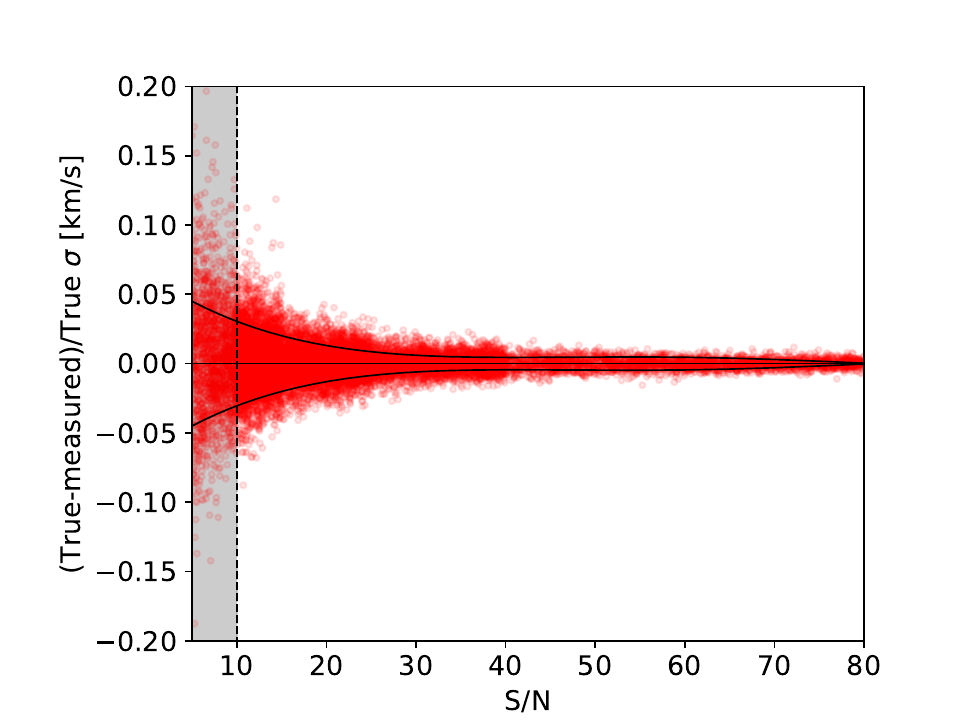}
  \includegraphics[width = \columnwidth]{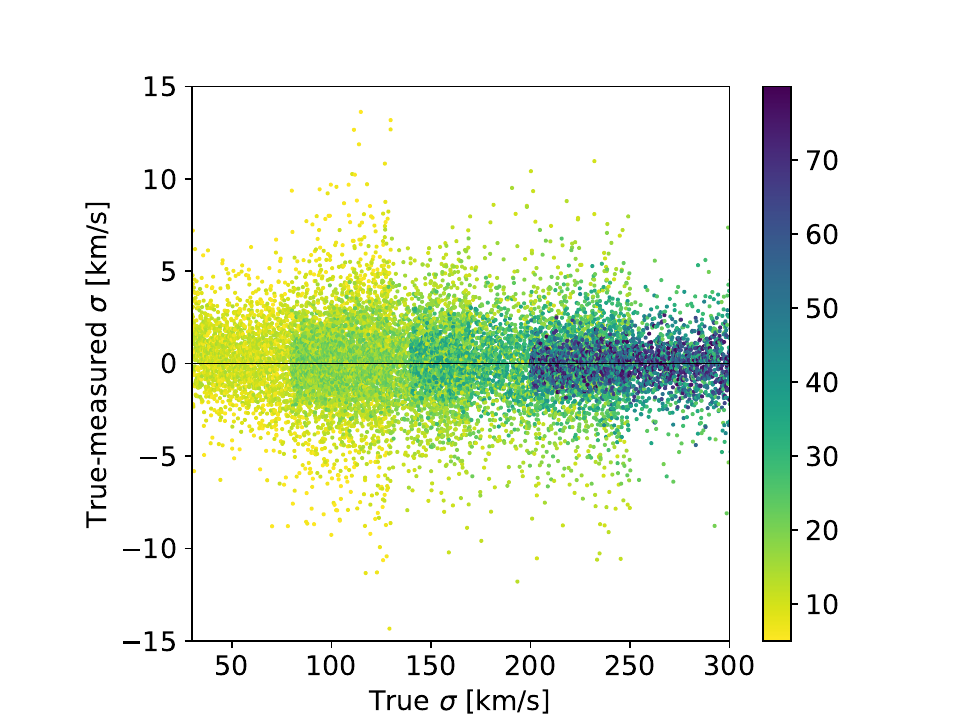}
  \caption{Velocity dispersion measurements for simulated spectra. Left panel: Relative offset of the recovered velocity dispersion with respect to the true value as a function of the $S/N$. We show the median relative uncertainty from Eq. (\ref{errapp}) as a black line. Right panel: Absolute offset between the recovered and the true velocity dispersion values as a function of the input velocity dispersion. The points are colour-coded by the corresponding spectral $S/N$.}
  \label{fig:sims}
\end{figure*}

As a first test, we study the difference between the measured velocity dispersion $\sigma_m$ and the input value of $\sigma$. As is clear from the left panel of Fig. \ref{fig:sims}, the average offset is very close to zero, and on average $\approx 0.03 \, \mathrm{km \, s^{-1}}$ for the members with a reliable fit ($S/N>10$). Unlike \citet{bergamini19}, we find that we do not need to correct for any bias to the value of $\sigma$, even in the range $\sigma<100 \, \mathrm{km \, s^{-1}}$.

As a following step, we tested the impact of the spectral noise on the uncertainty on the velocity dispersion value. Even for $S/N>10$, the noise determines random offsets between the recovered and true values of the velocity dispersion. During the LOSVD fit, \texttt{pPXF} determines the statistical uncertainty on the best-fit value of $\sigma$ required to fit the stellar templates to the observed spectrum. As found by \citet{cappellari04}, this method can yield a significantly underestimated uncertainty value for noisy spectra and/or lower $\sigma$ values. As detailed in Sect. \ref{s3}, we  increased this value by adding in quadrature another contribution from the random offsets of $\sigma$ determined by the spectral noise. To estimate this contribution, we used our simulated spectra to study the relative offsets between the measured and observed values of $\sigma$ as a function of $S/N$. We choose a set of $S/N$ bins and take the standard deviation of the difference between the measured and input velocity dispersion in each bin, which we indicate as $|\delta \sigma_\mathrm{sim}|$. As shown in Fig. \ref{fig:sims}, the relative offset $|\delta \sigma|/\sigma$ is typically smaller than $5\%$ for $S/N>10$ and goes to zero for $S/N\approx80$. A polynomial fit of the relative uncertainty $\delta \sigma/\sigma$ as a function of the value of $S/N$ gives the following relation
\begin{multline}\label{errapp}
    |\delta \sigma_\mathrm{sim}|/\sigma = 0.0705-5.1 \times 10^{-3}\, S/N+1.38\times10^{-4}\, (S/N)^2 \\-1.57\times 10^{-6}\, (S/N)^3+6.29\times 10^{-9}\, (S/N)^4,
\end{multline}
where $\sigma$ is the measured value of the LOS velocity dispersion provided by \texttt{pPXF}. In Sect. \ref{s3} we detail how we take advantage of this relation to estimate the total uncertainty on the values of the stellar velocity dispersion we report in Appendix \ref{appc}.

\begin{table}
\caption{Number of simulated spectra $N$ of cluster members in four bins of velocity dispersion,  value intervals for $\sigma$, and $S/N$ of the simulated spectra.}             % title of Table
\centering                          % used for centering table
\begin{tabular}{c c c}        % centered columns (4 columns)
\hline \noalign{\smallskip}                % inserts double horizontal lines
Number of spectra & $\sigma\,[\mathrm{km \, s^{-1}}]$ & $S/N$ \\    % table heading 
\noalign{\smallskip} \hline  \noalign{\smallskip}                      % inserts single horizontal line
            4000 & $[30-130]$ & $[5-15]$  \\
            \rule{0pt}{2.5ex}
            4000 & $[80-170]$ & $[10-25]$\\
            \rule{0pt}{2.5ex}
            4000 & $[140-250]$ & $[10-40]$     \\
            \rule{0pt}{2.5ex}
            4000 & $[200-350]$ & $[20-80]$    \\
 \noalign{\smallskip} \hline                                   %inserts single line
\end{tabular}\label{simpar}
\end{table}

Finally, we test the impact of the optimisation set-up chosen for this work and presented in Sect. \ref{s3}. We confirm that the accuracy in recovering the value of the LOSVD of the simulated spectra presented in Table \ref{simpar} is not affected by the choice of a different degree for the additive polynomials adopted for the spectral fit nor improved by the inclusion of multiplicative polynomials. We verify that masking some pixels of the wavelength range, which is necessary to avoid including contaminated regions of the observed spectra, does not bias the measured values of the velocity dispersion. This is also the case when some absorption lines are masked, as is often the case (see Fig. \ref{showfit}), although we notice an increase in the scatter on the recovered values when either the CaII H or K lines are not included in the spectral fit.

\section{Catalogues of structural parameters}
\label{appendix:cat_description}

The catalogues of structural parameters for the four clusters A2744, AS1063, M0416, and M1149 are available at the CDS via anonymous ftp to cdsarc.cds.unistra.fr or via https://cdsarc.cds.unistra.fr/viz-bin/cat/J/A+A/709/A254. The table column names (in typewritten font) and their descriptions are as follows:
\begin{itemize}
\item $\tt{ID}$: ID in serial order.
\item  $\tt{RA}$: right ascension in degrees.
\item  $\tt{DEC}$: declination in degrees.
\item  $\tt{MAG\_x}$: total magnitude in the x-band.
\item  $\tt{MAG\_ERR\_x}$: total magnitude error in the x-band.
\item  $\tt{RE\_x}$: circularised effective radius in kiloparsecs in the x-band.
\item  $\tt{RE\_ERR\_x}$: circularised effective radius error in kiloparsecs in the x-band.
\item  $\tt{MU\_x}$: average surface brightness within the effective radius in $\mathrm{mag\ arcsec^{-2}}$ in the x-band.
\item  $\tt{MU\_ERR\_x}$: average surface brightness within the effective radius error in $\mathrm{mag\ arcsec^{-2}}$ in the x-band.
\item  $\tt{N\_x}$: S\'ersic index in the x-band.
\item  $\tt{N\_ERR\_x}$: S\'ersic index error in the x-band.
\item  $\tt{AR\_x}$: axis ratio in the x-band.
\item  $\tt{AR\_ERR\_x}$: axis ratio error in the x-band.
\item  $\tt{PA\_x}$: position angle in degrees in the x-band.
\item  $\tt{PA\_ERR\_x}$: position angle error in degrees in the x-band.
\end{itemize}

$\tt{MAG\_x}$,   $\tt{N\_x}$,  $\tt{AR\_x}$,  $\tt{PA\_x}$ and their respective errors are the final structural parameters obtained from the fit on the full images. The weighted means of the semi-major axes in pixels and of the axis ratios in the $\mathrm{x}$-band ($\tt{r_{e}\_x}$ and $\tt{AR\_x}$, and their errors $\tt{r_{e}\_ERR\_x}$ and $\tt{AR\_ERR\_x}$) are used to compute the circularised effective radii in kiloparsecs, 
\begin{equation}
\tt{RE\_x} = \tt{r_{e}\_x} \times \sqrt{\tt{AR\_x}} \times \mathrm{pixel\_scale} \times \mathrm{kpc\_per\_arcsec} \ ,
\end{equation}
and their errors $\tt{RE\_ERR\_x}$. These are combined with the magnitudes in the $\mathrm{x}$-band $\tt{MAG\_x}$ and their errors $\tt{MAG\_ERR\_x}$ to compute the average surface brightnesses within that radius,
\begin{eqnarray}
\label{weighted_average_formula}
\nonumber \tt{MU\_x} &=&  \tt{MAG\_x} + 2.5 \log{(2 \pi)} + 5 \log{(\tt{RE\_x}}  / \mathrm{kpc\_per\_arcsec}) \\
&-& 10 \log{(1 + z_{\mathrm{cluster}})} \ ,
\end{eqnarray}
and their errors $\tt{MU\_ERR\_x}$.

\onecolumn
\section{Catalogues of measured velocity dispersion values}\label{appc}
In this appendix we present the final kinematic catalogues for the four clusters. We include 76 members of A2744, 34 of AS1063, 52 of M0416, and 51 of M1149. The procedure for the kinematic measurements is described in Sect. \ref{s3}. The IDs and coordinates of the members follow those reported in Appendix \ref{appendix:cat_description}.

\begin{longtable}{c c c c c c c}
\caption{\label{tab:table2744} Velocity dispersion catalogue for A2744.}\\
\hline\hline
ID & R.A. & Dec & ID\textsubscript{MUSE} & $\sigma \, [\mathrm{km \, s^{-1}}]$ & $\delta \sigma \, [\mathrm{km \, s^{-1}}]$ & $S/N$ \\  & (J2000) & (J2000) & & & & \\
\hline
\endfirsthead
\caption{continued.}\\
\hline\hline
ID & R.A. & Dec & ID\textsubscript{MUSE} & $\sigma \, [\mathrm{km \, s^{-1}}]$ & $\delta \sigma \, [\mathrm{km \, s^{-1}}]$ & $S/N$ \\  & (J2000) & (J2000) & & & & \\
\hline
\endhead
\hline
\endfoot
2 & 3.60136 & $-$30.41536 & 32547 & 88.2 & 1.8 & 51.0 \\
4 & 3.58571 & $-$30.41397 & 32768 & 57.5 & 3.8 & 23.9 \\
5 & 3.56958 & $-$30.41216 & 33328 & 91.7 & 2.5 & 40.0 \\
6 & 3.57967 & $-$30.40919 & 34423 & 181.0 & 1.2 & 102.0 \\
\textbf{8} & 3.59203 & $-$30.40573 & 36034 & 287.2 & 1.7 & 123.0 \\
9 & 3.57395 & $-$30.40883 & 35061 & 178.2 & 1.9 & 67.5 \\
11 & 3.58882 & $-$30.41072 & 33540 & 106.6 & 2.5 & 36.7 \\
15 & 3.58251 & $-$30.40999 & 33699 & 117.6 & 3.4 & 28.5 \\
17 & 3.58913 & $-$30.40957 & 33910 & 110.6 & 2.0 & 43.9 \\
19 & 3.58972 & $-$30.41023 & 33410 & 79.8 & 6.5 & 13.1 \\
22 & 3.58383 & $-$30.40951 & 33671 & 70.6 & 5.6 & 11.5 \\
23 & 3.59172 & $-$30.40781 & 34556 & 185.5 & 3.2 & 44.9 \\
26 & 3.59786 & $-$30.40555 & 21367 & 77.7 & 2.6 & 36.8 \\
\textbf{29} & 3.58625 & $-$30.40016 & 37824 & 304.0 & 1.9 & 114.5 \\
31 & 3.57340 & $-$30.40746 & 34538 & 33.3 & 7.5 & 10.9 \\
33 & 3.59591 & $-$30.40621 & 35339 & 135.3 & 2.0 & 46.9 \\
34 & 3.59250 & $-$30.40462 & 36210 & 222.2 & 1.4 & 116.3 \\
36 & 3.57851 & $-$30.40338 & 36527 & 214.8 & 2.0 & 107.7 \\
37 & 3.59028 & $-$30.40740 & 34439 & 97.3 & 3.4 & 22.2 \\
39 & 3.59896 & $-$30.39753 & 39072 & 98.0 & 1.7 & 48.6 \\
40 & 3.58704 & $-$30.40495 & 35693 & 152.7 & 2.5 & 45.8 \\
45 & 3.60527 & $-$30.40081 & 37068 & 229.9 & 3.5 & 51.3 \\
49 & 3.58794 & $-$30.40085 & 36892 & 96.3 & 1.5 & 63.1 \\
50 & 3.60528 & $-$30.40293 & 36220 & 73.5 & 2.6 & 24.7 \\
53 & 3.58438 & $-$30.40289 & 36043 & 85.9 & 3.0 & 24.2 \\
57 & 3.59483 & $-$30.40213 & 36298 & 58.1 & 3.7 & 18.0 \\
60 & 3.60435 & $-$30.40012 & 37344 & 158.7 & 1.9 & 61.6 \\
61 & 3.59300 & $-$30.39933 & 37947 & 153.5 & 1.1 & 95.4 \\
64 & 3.57859 & $-$30.39911 & 37609 & 70.1 & 4.2 & 24.4 \\
65 & 3.58766 & $-$30.39640 & 39382 & 218.3 & 1.3 & 116.8 \\
66 & 3.59446 & $-$30.40035 & 37229 & 174.0 & 2.5 & 48.4 \\
67 & 3.59480 & $-$30.39165 & 40689 & 212.8 & 1.1 & 132.0 \\
69 & 3.57281 & $-$30.40053 & 36953 & 94.4 & 4.5 & 15.3 \\
70 & 3.59676 & $-$30.40051 & 36849 & 68.7 & 4.3 & 14.3 \\
72 & 3.58216 & $-$30.39857 & 38117 & 162.8 & 1.4 & 75.5 \\
73 & 3.57490 & $-$30.39838 & 38067 & 180.9 & 1.7 & 76.0 \\
75 & 3.57907 & $-$30.40009 & 36843 & 58.6 & 5.8 & 22.0 \\
78 & 3.58398 & $-$30.39926 & 37230 & 86.9 & 2.3 & 51.2 \\
81 & 3.57886 & $-$30.39711 & 38729 & 120.0 & 1.8 & 54.2 \\
84 & 3.58292 & $-$30.39970 & 36982 & 35.0 & 12.2 & 22.5 \\
85 & 3.58838 & $-$30.39836 & 38010 & 157.2 & 1.5 & 81.6 \\
86 & 3.60271 & $-$30.39757 & 37825 & 64.8 & 4.8 & 15.1 \\
92 & 3.58552 & $-$30.39715 & 38275 & 61.0 & 2.3 & 32.0 \\
94 & 3.58868 & $-$30.39608 & 38930 & 148.2 & 1.0 & 105.3 \\
95 & 3.60220 & $-$30.39699 & 38143 & 84.2 & 4.1 & 18.8 \\
96 & 3.58539 & $-$30.39428 & 40059 & 210.6 & 2.5 & 78.3 \\
97 & 3.59287 & $-$30.39635 & 38252 & 63.4 & 4.5 & 13.6 \\
98 & 3.57895 & $-$30.39412 & 39646 & 131.1 & 2.6 & 41.9 \\
99 & 3.58815 & $-$30.39507 & 39428 & 150.4 & 1.1 & 99.0 \\
102 & 3.60083 & $-$30.39490 & 39283 & 82.7 & 2.8 & 25.7 \\
104 & 3.58139 & $-$30.39393 & 39503 & 124.2 & 2.0 & 45.9 \\
108 & 3.58498 & $-$30.39288 & 39710 & 95.2 & 1.8 & 49.2 \\
113 & 3.57150 & $-$30.39043 & 40478 & 91.8 & 1.7 & 45.9 \\
116 & 3.58037 & $-$30.39220 & 39876 & 82.4 & 2.7 & 40.6 \\
118 & 3.58095 & $-$30.39081 & 40243 & 120.5 & 2.9 & 76.8 \\
119 & 3.58530 & $-$30.38756 & 41856 & 104.9 & 2.1 & 69.7 \\
120 & 3.59034 & $-$30.39094 & 40314 & 146.7 & 1.6 & 66.9 \\
121 & 3.58438 & $-$30.39175 & 39727 & 57.4 & 5.2 & 11.4 \\
124 & 3.57017 & $-$30.38645 & 41644 & 228.7 & 6.6 & 22.6 \\
125 & 3.57835 & $-$30.38946 & 40428 & 53.8 & 4.8 & 13.3 \\
127 & 3.59471 & $-$30.38912 & 42443 & 90.3 & 1.4 & 49.2 \\
130 & 3.59317 & $-$30.39035 & 40239 & 69.1 & 6.6 & 18.0 \\
136 & 3.58919 & $-$30.38740 & 41950 & 188.5 & 1.6 & 79.0 \\
137 & 3.59551 & $-$30.38868 & 42269 & 108.9 & 2.2 & 36.6 \\
138 & 3.59877 & $-$30.38802 & 42149 & 121.8 & 2.0 & 41.1 \\
139 & 3.58372 & $-$30.38467 & 41303 & 59.5 & 3.2 & 38.0 \\
141 & 3.59255 & $-$30.38531 & 41418 & 189.6 & 2.9 & 46.4 \\
144 & 3.57834 & $-$30.38710 & 42195 & 55.0 & 3.6 & 35.1 \\
145 & 3.58814 & $-$30.38501 & 41363 & 143.7 & 2.2 & 46.4 \\
148 & 3.59347 & $-$30.38760 & 42079 & 65.0 & 10.4 & 10.9 \\
149 & 3.59328 & $-$30.38438 & 41259 & 125.0 & 3.6 & 58.9 \\
150 & 3.60543 & $-$30.38484 & 41440 & 179.3 & 2.1 & 56.0 \\
154 & 3.60439 & $-$30.38496 & 41908 & 66.7 & 6.2 & 29.1 \\
156 & 3.59028 & $-$30.38269 & 40884 & 118.9 & 2.1 & 39.8 \\
161 & 3.58804 & $-$30.38256 & 40832 & 82.0 & 3.2 & 24.1 \\
166 & 3.57847 & $-$30.38132 & 44545 & 89.9 & 5.2 & 15.2 \\

\end{longtable}
\tablefoot{
We identify the 76 galaxies included in this catalogue with photometric catalogue ID (see Appendix \ref{appendix:cat_description}), RA, Dec, and MUSE ID. We report the measured stellar velocity dispersion $\sigma$ and its uncertainty $\delta \sigma$, both in $\mathrm{km\,s^{-1}}$, and the spectral $S/N$. The BCGs are marked in boldface.}

\begin{longtable}{c c c c c c c}
\caption{\label{tab:table2248} Velocity dispersion catalogue for AS1063.}\\
\hline\hline
ID & R.A. & Dec & ID\textsubscript{MUSE} & $\sigma \, [\mathrm{km \, s^{-1}}]$ & $\delta \sigma \, [\mathrm{km \, s^{-1}}]$ & $S/N$ \\  & (J2000) & (J2000) & & & & \\
\hline
\endfirsthead
\caption{continued.}\\
\hline\hline
ID & R.A. & Dec & ID\textsubscript{MUSE} & $\sigma \, [\mathrm{km \, s^{-1}}]$ & $\delta \sigma \, [\mathrm{km \, s^{-1}}]$ & $S/N$ \\  & (J2000) & (J2000) & & & & \\
\hline
\endhead
\hline
\endfoot
3 & 342.18449 & $-$44.54318 & 35399 & 107.1 & 5.3 & 16.1 \\
6 & 342.17176 & $-$44.54054 & 35670 & 115.9 & 4.3 & 20.4 \\
7 & 342.18666 & $-$44.54076 & 35719 & 131.5 & 6.2 & 17.1 \\
8 & 342.18893 & $-$44.54039 & 35720 & 120.4 & 8.6 & 13.1 \\
9 & 342.18205 & $-$44.54034 & 35775 & 99.5 & 7.3 & 15.0 \\
14 & 342.16046 & $-$44.53895 & 35984 & 65.1 & 11.8 & 10.3 \\
18 & 342.18438 & $-$44.53619 & 36392 & 136.2 & 3.4 & 30.0 \\
19 & 342.17547 & $-$44.53539 & 35779 & 133.1 & 4.6 & 41.2 \\
22 & 342.18693 & $-$44.53537 & 36510 & 91.0 & 5.0 & 18.3 \\
24 & 342.16985 & $-$44.53554 & 36486 & 144.2 & 4.7 & 22.3 \\
30 & 342.19134 & $-$44.53432 & 36649 & 172.2 & 6.0 & 20.9 \\
33 & 342.17492 & $-$44.53413 & 36679 & 124.8 & 6.2 & 15.3 \\
34 & 342.17903 & $-$44.53278 & 36880 & 251.2 & 5.9 & 33.3 \\
\textbf{35} & 342.18322 & $-$44.53088 & 33294 & 332.3 & 2.7 & 99.5 \\
37 & 342.19156 & $-$44.53334 & 36793 & 110.5 & 7.3 & 12.8 \\
42 & 342.17793 & $-$44.53239 & 36936 & 156.7 & 6.1 & 22.7 \\
47 & 342.18109 & $-$44.52920 & $-$80 & 81.1 & 10.0 & 10.6 \\
51 & 342.18916 & $-$44.52953 & 37335 & 164.6 & 4.1 & 37.8 \\
54 & 342.18813 & $-$44.52595 & 37522 & 286.9 & 4.7 & 46.3 \\
55 & 342.18678 & $-$44.52781 & 37550 & 114.7 & 4.0 & 29.7 \\
56 & 342.19551 & $-$44.52599 & 37572 & 238.2 & 2.6 & 71.8 \\
57 & 342.19330 & $-$44.52643 & 37760 & 175.4 & 3.6 & 37.2 \\
60 & 342.18405 & $-$44.52693 & 37685 & 128.7 & 3.1 & 36.1 \\
63 & 342.20029 & $-$44.52520 & 37949 & 134.0 & 3.3 & 31.2 \\
64 & 342.18255 & $-$44.52687 & 37695 & 92.8 & 7.0 & 13.7 \\
69 & 342.17891 & $-$44.52472 & 38024 & 66.0 & 6.2 & 12.3 \\
70 & 342.17962 & $-$44.52305 & 38261 & 80.8 & 5.7 & 28.1 \\
73 & 342.19329 & $-$44.51782 & 38895 & 290.7 & 4.6 & 52.6 \\
79 & 342.18543 & $-$44.51863 & 38782 & 250.4 & 3.4 & 56.9 \\
83 & 342.19294 & $-$44.52207 & 38389 & 52.4 & 14.1 & 10.6 \\
84 & 342.19727 & $-$44.52327 & 38246 & 165.0 & 3.8 & 35.7 \\
85 & 342.18665 & $-$44.52247 & 38339 & 136.9 & 4.2 & 24.2 \\
87 & 342.17795 & $-$44.52406 & 38129 & 90.0 & 7.2 & 12.0 \\
92 & 342.20418 & $-$44.52524 & 37939 & 141.8 & 3.4 & 30.7 \\

\end{longtable}
\tablefoot{
We identify the 34 galaxies included in this catalogue with photometric catalogue ID (see Appendix \ref{appendix:cat_description}), RA, Dec, and MUSE ID. We report the measured stellar velocity dispersion $\sigma$ and its uncertainty $\delta \sigma$, both in $\mathrm{km\,s^{-1}}$, and the spectral $S/N$. The BCG is marked in boldface.}

\begin{longtable}{c c c c c c c}
\caption{\label{tab:table0416} Velocity dispersion catalogue for M0416.}\\
\hline\hline
ID & R.A. & Dec & ID\textsubscript{MUSE} & $\sigma \, [\mathrm{km \, s^{-1}}]$ & $\delta \sigma \, [\mathrm{km \, s^{-1}}]$ & $S/N$ \\  & (J2000) & (J2000) & & & & \\
\hline
\endfirsthead
\caption{continued.}\\
\hline\hline
ID & R.A. & Dec & ID\textsubscript{MUSE} & $\sigma \, [\mathrm{km \, s^{-1}}]$ & $\delta \sigma \, [\mathrm{km \, s^{-1}}]$ & $S/N$ \\  & (J2000) & (J2000) & & & & \\
\hline
\endhead
\hline
\endfoot
\textbf{1} & 64.03810 & $-$24.06749 & 79744 & 275.6 & 1.5 & 164.6 \\
2 & 64.04486 & $-$24.07351 & 80726 & 209.3 & 1.4 & 149.3 \\
3 & 64.03471 & $-$24.07160 & 81285 & 232.6 & 3.7 & 51.3 \\
4 & 64.03823 & $-$24.07176 & 81838 & 191.4 & 2.1 & 77.4 \\
5 & 64.04276 & $-$24.07303 & 81883 & 139.7 & 1.6 & 78.3 \\
6 & 64.03268 & $-$24.07015 & 82042 & 116.8 & 3.8 & 28.1 \\
8 & 64.04636 & $-$24.06706 & 83064 & 186.7 & 1.6 & 115.5 \\
9 & 64.02831 & $-$24.07227 & 82442 & 175.5 & 3.3 & 49.2 \\
10 & 64.04245 & $-$24.07465 & 82012 & 48.7 & 6.4 & 15.8 \\
12 & 64.04131 & $-$24.07136 & 82612 & 194.3 & 2.0 & 107.9 \\
13 & 64.03679 & $-$24.07226 & 81782 & 88.8 & 6.6 & 13.6 \\
15 & 64.04010 & $-$24.06591 & 82063 & 185.7 & 1.2 & 132.6 \\
16 & 64.04370 & $-$24.07337 & 82585 & 72.8 & 6.8 & 19.9 \\
18 & 64.03945 & $-$24.06933 & 82766 & 168.2 & 1.2 & 113.8 \\
19 & 64.04516 & $-$24.07145 & 82961 & 90.3 & 4.1 & 22.2 \\
20 & 64.03473 & $-$24.06980 & 82611 & 134.6 & 3.1 & 40.1 \\
23 & 64.03878 & $-$24.07072 & 82024 & 90.1 & 1.7 & 56.7 \\
24 & 64.03344 & $-$24.06907 & 82598 & 109.4 & 5.4 & 20.5 \\
26 & 64.04269 & $-$24.06513 & 83948 & 107.7 & 2.1 & 56.7 \\
29 & 64.03519 & $-$24.06874 & 83541 & 63.9 & 4.0 & 24.4 \\
30 & 64.04251 & $-$24.06923 & 83042 & 208.7 & 1.5 & 115.0 \\
32 & 64.04418 & $-$24.06872 & 83375 & 115.1 & 2.8 & 92.8 \\
34 & 64.04481 & $-$24.06708 & 84025 & 181.5 & 1.9 & 109.6 \\
35 & 64.03574 & $-$24.06968 & 82863 & 215.9 & 6.1 & 36.1 \\
38 & 64.03480 & $-$24.06282 & 85939 & 68.5 & 7.3 & 12.4 \\
40 & 64.03445 & $-$24.06797 & 84115 & 55.6 & 8.7 & 11.8 \\
45 & 64.03872 & $-$24.06610 & 84536 & 135.2 & 2.5 & 69.1 \\
48 & 64.04612 & $-$24.06395 & 85357 & 97.1 & 2.4 & 41.0 \\
50 & 64.04253 & $-$24.06324 & 84395 & 211.4 & 1.7 & 106.1 \\
51 & 64.04285 & $-$24.06222 & 86038 & 96.2 & 1.9 & 46.6 \\
53 & 64.03852 & $-$24.06207 & 85770 & 126.2 & 1.5 & 79.9 \\
54 & 64.04519 & $-$24.06213 & 84757 & 211.0 & 1.6 & 120.5 \\
55 & 64.04088 & $-$24.06035 & 86229 & 114.2 & 1.5 & 70.9 \\
56 & 64.05061 & $-$24.06462 & 82505 & 51.3 & 8.3 & 11.1 \\
57 & 64.03816 & $-$24.06312 & 85812 & 69.8 & 2.4 & 31.2 \\
60 & 64.05262 & $-$24.06571 & 85042 & 53.7 & 6.0 & 14.4 \\
66 & 64.03687 & $-$24.08067 & 77463 & 222.6 & 1.9 & 105.7 \\
68 & 64.02845 & $-$24.08909 & -144 & 78.2 & 5.1 & 16.3 \\
70 & 64.02734 & $-$24.08660 & 78004 & 89.7 & 8.9 & 25.3 \\
71 & 64.02547 & $-$24.08510 & 77317 & 153.5 & 4.1 & 33.5 \\
\textbf{73} & 64.03197 & $-$24.07743 & 77342 & 286.4 & 2.1 & 134.1 \\
75 & 64.03932 & $-$24.07686 & 79058 & 193.0 & 3.2 & 51.2 \\
77 & 64.02221 & $-$24.08526 & 78230 & 70.3 & 4.6 & 18.1 \\
83 & 64.02938 & $-$24.07901 & 79978 & 131.8 & 3.0 & 47.2 \\
89 & 64.03384 & $-$24.08056 & 79957 & 64.9 & 6.8 & 14.1 \\
92 & 64.03820 & $-$24.07897 & 79898 & 65.2 & 4.4 & 15.8 \\
93 & 64.03521 & $-$24.07724 & 79996 & 116.6 & 2.4 & 52.8 \\
97 & 64.02198 & $-$24.07786 & 80466 & 198.7 & 4.4 & 42.0 \\
102 & 64.03441 & $-$24.07536 & 81093 & 119.6 & 16.8 & 11.6 \\
103 & 64.02671 & $-$24.07639 & 80895 & 90.5 & 5.4 & 20.3 \\
107 & 64.03378 & $-$24.07617 & 81125 & 145.6 & 7.6 & 19.9 \\
109 & 64.02888 & $-$24.07506 & 81561 & 98.5 & 6.0 & 19.6 \\

\end{longtable}
\tablefoot{
We identify the 52 galaxies included in this catalogue with photometric catalogue ID (see Appendix \ref{appendix:cat_description}), RA, Dec, and MUSE ID. We report the measured stellar velocity dispersion $\sigma$ and its uncertainty $\delta \sigma$, both in $\mathrm{km\,s^{-1}}$, and the spectral $S/N$. The BCGs are marked in boldface.}

\begin{longtable}{c c c c c c c}
\caption{\label{tab:table1149} Velocity dispersion catalogue for M1149.}\\
\hline\hline
ID & R.A. & Dec & ID\textsubscript{MUSE} & $\sigma \, [\mathrm{km \, s^{-1}}]$ & $\delta \sigma \, [\mathrm{km \, s^{-1}}]$ & $S/N$ \\  & (J2000) & (J2000) & & & & \\
\hline
\endfirsthead
\caption{continued.}\\
\hline\hline
ID & R.A. & Dec & ID\textsubscript{MUSE} & $\sigma \, [\mathrm{km \, s^{-1}}]$ & $\delta \sigma \, [\mathrm{km \, s^{-1}}]$ & $S/N$ \\  & (J2000) & (J2000) & & & & \\
\hline
\endhead
\hline
\endfoot
2 & 177.40645 & 22.38958 & 3908 & 266.8 & 3.5 & 54.3 \\
3 & 177.40369 & 22.38911 & 4772 & 117.7 & 6.7 & 15.3 \\
4 & 177.39168 & 22.39062 & 5057 & 125.9 & 3.7 & 28.9 \\
5 & 177.39855 & 22.38979 & 5353 & 108.9 & 7.7 & 11.0 \\
8 & 177.39869 & 22.39230 & 4814 & 129.5 & 8.9 & 19.9 \\
9 & 177.40366 & 22.39194 & 5589 & 205.2 & 9.2 & 27.2 \\
10 & 177.39266 & 22.39273 & 5503 & 205.4 & 3.7 & 38.7 \\
12 & 177.39123 & 22.39272 & 4969 & 130.0 & 8.2 & 18.1 \\
16 & 177.38945 & 22.39391 & 5388 & 199.4 & 7.0 & 22.8 \\
17 & 177.39583 & 22.39350 & 5701 & 160.4 & 11.9 & 24.0 \\
18 & 177.39288 & 22.39710 & 6445 & 136.1 & 5.5 & 19.7 \\
20 & 177.39269 & 22.39436 & 6124 & 144.0 & 8.0 & 13.6 \\
21 & 177.40358 & 22.39638 & 6006 & 250.4 & 4.0 & 45.8 \\
22 & 177.39778 & 22.39545 & 5810 & 236.7 & 6.6 & 29.0 \\
24 & 177.39278 & 22.39809 & 6132 & 166.3 & 3.7 & 30.9 \\
\textbf{26} & 177.39875 & 22.39854 & 5480 & 271.2 & 4.4 & 47.5 \\
27 & 177.40262 & 22.39619 & 5620 & 140.0 & 7.6 & 15.4 \\
30 & 177.39983 & 22.39726 & 6338 & 172.8 & 5.6 & 28.2 \\
32 & 177.40103 & 22.39788 & 6340 & 303.8 & 7.1 & 32.5 \\
35 & 177.40544 & 22.39787 & 6670 & 187.0 & 4.3 & 26.9 \\
38 & 177.40752 & 22.40305 & 7342 & 299.4 & 5.0 & 44.5 \\
42 & 177.40226 & 22.39975 & 7170 & 166.9 & 6.5 & 18.8 \\
44 & 177.40120 & 22.40033 & 7207 & 188.9 & 4.1 & 31.1 \\
45 & 177.39752 & 22.39954 & 6752 & 191.1 & 9.8 & 19.2 \\
46 & 177.39965 & 22.39961 & 7070 & 74.8 & 7.3 & 11.5 \\
47 & 177.39796 & 22.40105 & 6697 & 182.5 & 3.8 & 36.6 \\
49 & 177.39097 & 22.40168 & 7566 & 106.4 & 3.3 & 25.1 \\
55 & 177.39886 & 22.40181 & 7463 & 147.4 & 7.7 & 13.2 \\
61 & 177.40372 & 22.40458 & 10247 & 193.5 & 5.0 & 26.6 \\
62 & 177.39015 & 22.40390 & 13114 & 174.0 & 9.8 & 13.6 \\
63 & 177.39109 & 22.40491 & 10466 & 254.7 & 4.5 & 40.9 \\
66 & 177.39183 & 22.40529 & 10542 & 225.6 & 5.4 & 29.6 \\
169 & 177.39815 & 22.40741 & 12242 & 212.0 & 5.2 & 31.3 \\
177 & 177.39900 & 22.41839 & 10144 & 116.1 & 4.8 & 19.0 \\
197 & 177.39826 & 22.41079 & 11671 & 170.5 & 3.3 & 36.0 \\
220 & 177.38947 & 22.40638 & 12900 & 121.7 & 4.5 & 22.5 \\
226 & 177.38811 & 22.40841 & 12169 & 218.8 & 4.3 & 36.0 \\
234 & 177.38390 & 22.41028 & 11759 & 136.6 & 4.3 & 24.5 \\
246 & 177.38354 & 22.41494 & 10943 & 56.1 & 5.7 & 23.1 \\
247 & 177.39524 & 22.41493 & 10946 & 122.3 & 5.7 & 17.1 \\
248 & 177.38887 & 22.41415 & 11033 & 111.6 & 7.0 & 14.6 \\
250 & 177.39772 & 22.41521 & 10838 & 133.7 & 12.6 & 10.4 \\
252 & 177.39383 & 22.41169 & 11372 & 182.4 & 3.6 & 38.7 \\
254 & 177.39472 & 22.41441 & 11038 & 106.5 & 5.1 & 17.4 \\
255 & 177.38811 & 22.41376 & 11009 & 207.7 & 6.4 & 23.1 \\
258 & 177.39659 & 22.41333 & 11151 & 177.4 & 3.5 & 33.8 \\
262 & 177.38275 & 22.41310 & 10891 & 107.6 & 7.0 & 12.9 \\
268 & 177.39703 & 22.41156 & 11618 & 153.8 & 7.8 & 13.9 \\
274 & 177.39405 & 22.40844 & 12377 & 132.1 & 5.9 & 18.8 \\
276 & 177.39427 & 22.40756 & 12443 & 93.8 & 7.2 & 16.2 \\
279 & 177.39683 & 22.41746 & 10217 & 191.6 & 4.6 & 29.5 \\
\end{longtable}
\tablefoot{
We identify the 51 galaxies included in this catalogue with photometric catalogue ID (see Appendix \ref{appendix:cat_description}), RA, Dec, and MUSE ID. We report the measured stellar velocity dispersion $\sigma$ and its uncertainty $\delta \sigma$, both in $\mathrm{km\,s^{-1}}$, and the spectral $S/N$. The BCG is marked in boldface.}

\end{appendix}

\end{document}